\documentclass[manuscript, screen, nonacm]{acmart}

\usepackage{graphicx}
\usepackage{setspace}               % for LINE SPACING
\usepackage{multicol}               % for MULTICOLUMNS
\usepackage{xcolor}
\usepackage{caption}
\usepackage{subcaption}
\usepackage{csquotes}

\usepackage[utf8]{inputenc}
\usepackage{url}
\usepackage{hyperref}
\usepackage{amsfonts}
\usepackage{float} %For [H] in figures
\usepackage{listings}
\usepackage{color}
\usepackage{fancyhdr} %For headers http://tex.stackexchange.com/questions/118647/line-under-chapter-name-at-top-of-the-page
\usepackage{bibentry} 
\usepackage{enumitem}
\usepackage{multirow}
\usepackage{geometry}
\usepackage{dirtytalk}
\usepackage{tabularx}
\usepackage{balance}

\usepackage{array,ragged2e,hyphenat} % in the preamble
\newcolumntype{L}[1]{>{\RaggedRight\arraybackslash\nohyphens}p{#1}}

\usepackage[english]{babel}
\usepackage{soul}

\definecolor{edit}{HTML}{000000}

\usepackage[most]{tcolorbox} % Load tcolorbox with additional libraries
\newtcolorbox{boxA}{
    boxrule = 1pt,
    colframe = black % frame color
}

\usepackage{listings}
\lstdefinestyle{mystyle}{
    basicstyle=\ttfamily\footnotesize,
    breakatwhitespace=false,         
    breaklines=true,                 
    captionpos=b,                    
    keepspaces=true,                 
    numbers=left,                    
    numbersep=5pt,                  
    showspaces=false,                
    showstringspaces=false,
    showtabs=false,                  
    tabsize=2
}
\AtBeginDocument{%
  }

\setcopyright{acmlicensed}
\acmConference[DIS '26]{ACM Designing Interactive Systems}{June 13-17, 2026}{National University of Singapore, Singapore}
\acmBooktitle{ACM Designing Interactive Systems, June 13-17, 2026, National University of Singapore, Singapore}

\acmISBN{978-1-4503-XXXX-X/2018/06}



\begin{document}

%\title[Lessening feelings of social judgement and apprehension via reasoning transparency]{"When the machine is thinking of me": Exploring user perceptions of chatbot thinking format when sharing personal problems}
\title[Exploring user perceptions of chatbot thinking format when sharing personal problems]{"When the machine is thinking of me": Exploring user perceptions of chatbot thinking format when sharing personal problems}
\title{How Does the Presence of a Chatbot's "Thinking" Affect User Perceptions When Sharing Personal Problems?}
\title{Watching AI Think: User Perceptions of Visible Thinking in Chatbots}

\author{Samuel Rhys Cox}
\email{srcox@cs.aau.dk}
\orcid{0000-0002-4558-6610}
\affiliation{%
  \institution{Aalborg University}
  \city{Aalborg}
  \country{Denmark}
}

\author{Jade Martin-Lise}
\email{jcpm@cs.aau.dk}
\orcid{0009-0002-9551-9608}
\affiliation{%
  \institution{Aalborg University}
  \city{Aalborg}
  \country{Denmark}
}

\author{Simo Hosio}
\email{simo.hosio@oulu.fi}
\orcid{0000-0002-9609-0965}
\affiliation{%
  \institution{University of Oulu}
  \city{Oulu}
  \country{Finland}
}

\author{Niels van Berkel}
\email{nielsvanberkel@cs.aau.dk}
\orcid{0000-0001-5106-7692}
\affiliation{%
  \institution{Aalborg University}
  \city{Aalborg}
  \country{Denmark}
}

%\renewcommand{\shortauthors}{Cox et al.}

\begin{abstract}
%People increasingly turn to chatbots such as ChatGPT in order to seek guidance for their personal problems.
People increasingly turn to conversational agents such as ChatGPT to seek guidance for their personal problems. 
As these systems grow in capability, many now display elements of ``\textit{thinking}''—short reflective statements that reveal a model's intentions or values before responding. 
While initially introduced to promote transparency, such visible thinking can also anthropomorphise the agent and shape user expectations. 
Yet little is known about how these displays affect user perceptions in help-seeking contexts.
We conducted a $3\times2$ mixed design experiment examining the impact of \textbf{Thinking Content} (\textsc{None}, \textsc{Emotionally-Supportive}, \textsc{Expertise-Supportive}) and \textbf{Conversation Context} (\textsc{Habit-Related} vs. \textsc{Feelings-Related} problems) on users' perceptions of empathy, warmth, competence, and engagement. 
Participants interacted with a chatbot that either showed no visible thinking or presented value-oriented reflections prior to its response.
Our findings contribute to understanding how \textit{thinking transparency} influences user experience in supportive dialogues, and offer implications for designing conversational agents that communicate intentions in sensitive, help-seeking scenarios.

%%%% Older version where was opening with "thinking" framing:
% As conversational agents (CAs) have expanded in adoption and capability, features such as CAs displaying chain-of-thought reasoning or ``thinking'' to users has become increasingly common.
% While such ``thinking'' was initially introduced to allow for decision-making transparency, recent commercially available CAs have allowed for the injection of anthropomorphisation and personality.
% In tandem to these capabilities, people are increasingly turning to CAs (such as ChatGPT) to seek guidance for their personal problems, and the effects of ``thinking'' when help-seeking is not known.
% Therefore, we conducted a 3 x 2 design investigating the impact of Thinking Content (3 levels) and Conversation Context (2 levels) in help-seeking situations.
% Specifically, participants spoke to (1) \textsc{None}: a CA with no visible thinking, (2) \textsc{Emotionally-Supportive}: a CA that presented thinking to users that focused on values of emotional provision, or (3) \textsc{Expertise-Supportive}: a CA that presented thinking to users that focused on values of expert provision.
% Across two different help-seeking scenarios (sharing \textsc{Habit-Related} or \textsc{Feelings-Related} personal problems), we explore the impact on user perceptions related to empathy, social perceptions, and engagement.
% Our results offer implications for the design of supportive technologies, as well as the use of CA thinking.
\end{abstract}

\maketitle

\section{Introduction}
The rapid development of large language models (LLMs) has led to widespread adoption of conversational AI across diverse tasks.
Among them is the use of LLMs, such as ChatGPT, for assisting with a wide range of personal problems and challenges~\cite{jung2025ve}.
Multiple studies and reports have described people's use of popular LLMs for cases such as mental well-being support ~\cite{kolding2025_ai_psychiatry, luo2025_digitaltherapist, song2025typingcureexperienceslarge}, companionship \cite{herbener2025lonely}, or as learning support \cite{rawas2024chatgpt}.
Prior work has investigated how conversational agents' social cues can be manipulated to make people feel more comfortable and supported (e.g.,~\cite{caforselfmanagement, chaves2021should, cox2025ephemerality}) when sharing personal problems, for example by providing details on the models' memory capabilities~\cite{cox2025ephemerality}.

Recent advances in Reasoning Language Models (RLMs) and Large Reasoning Models (LRMs) have popularised the idea of exposing a model's intermediate reasoning during response generation. 
These models are typically fine-tuned to produce \textit{reasoning traces}~\cite{illusion-of-thinking}, that is, step-by-step inferential explanations that reveal how an output is derived (e.g., OpenAI o-series~\cite{jaech2024_openai_o1}; DeepSeek R1~\cite{guo2025_deepseek_r1}).
In machine learning research, such traces serve as a means of algorithmic transparency, helping users or developers inspect the logical underpinning of an answer. 
The ``thinking'' visible in many contemporary conversational agents (CAs) differs markedly from these inferential outputs\footnote{Please note, the term `reasoning' is often used interchangeably for this functionality, and we stress that neither the terms `thinking' nor `reasoning' implies human-like logic or warrants anthropomorphisation of these computational systems.}. 
Commercial chatbots increasingly display reflective or value-oriented statements (e.g., ``\textit{I aim to be supportive and non-judgemental}'') that communicate intentions, goals, or social stance rather than deductive reasoning.
Our work builds on this emerging design space by examining thinking displays: short reflective utterances that signal a chatbot's guiding values or response strategy. 
Whereas RLM studies focus on faithfulness and reasoning accuracy, we investigate how visible thinking shapes perceptions of empathy, warmth, and competence. 
In doing so, we complement prior explainable AI and transparency research in HCI (e.g., conversational XAI~\cite{he2025conversationalxai,joshi2024explainability}) by highlighting the social and affective consequences of making an agent's ``thoughts'' explicit during help-seeking dialogues.

%However, the introduction of a CA's user-facing thinking presents a design space where the \todo{meta-cognition (?)} of a CA is...
%\todo{LLM thinking has been explored as a form of explanation}

In this article, we study the impact of such CA `thinking' presentations on users' perceptions in help-seeking scenarios.
%We follow a 3 $\times$ 2 mixed factorial design in which we manipulate the presentation of \textbf{Thinking Content}, in which we show thinking as either emotionally-supportive, expertise-supportive, or no thinking at all, and \textbf{Conversational Context}, asking participants to engage with a conversational agent in either a habit-centred or feelings-centred problem.
We followed a 3 $\times$ 2 mixed factorial design that manipulated \textbf{Thinking Content},
by presenting thinking as emotionally-supportive, expertise-supportive, or with no thinking content shown, 
% in which we showed thinking as either emotionally-supportive, expertise-supportive, or no thinking at all,
and \textbf{Conversational Context} by having participants discuss either a habit-centred or feelings-centred problem with the CA.
Our study investigated the impact of these two independent variables on participants' perceptions, such as the CA's perceived empathy and experience of the interaction.
Our results show that the visibility and framing of a CA's `thinking' shapes participant perceptions of empathy, warmth, and competence.
Specifically, \textsc{Emotionally-Supportive} thinking was seen as warm, empathic, and caring; while \textsc{Expertise-Supportive} thinking was seen as trustworthy and competent.
In contrast, the \textsc{None} condition (with no visible thinking) was described as unengaging, less warm, and produced perceptions of minimal effort or interest in the user.
Additionally, qualitative feedback found that \textit{Thinking Content} affected people's perceptions of chatbot capability, goodwill, and metaphors.
%This reasoning mode not only aims to provide higher-quality answers but also to give the user insights into how the generated answer was produced.
%This added level of transparency can help users identify where the LLM may have made incorrect assumptions or provide more insights into the LLM's underlying approach or intentions.

%While models can produce plausible rationales that \textit{justify} an answer, they do not reflect what actually drove the eventual output~\cite{turpin2023language}.

%Paragraph one (LLMs for solving people's problems, and also reasoning/thinking capabilities):
%\begin{itemize}
%    \item People increasing talk to large language models, such as ChatGPT, to help with personal problems~\cite{jung2025ve}.
%    \item For example, multiple studies and reports have described people's use of popular LLMs for cases such as mental well-being support, or...~\cite{}  ~\cite{kolding2025_ai_psychiatry} ~\cite{luo2025_digitaltherapist} ~\cite{song2025typingcureexperienceslarge}  
%    \item Twinned with this, LLMs have grown in capabilities with greater ``reasoning'' capabilities allowing for the generation of higher quality responses to user queries~\cite{}\footnote{Please note: \todo{potential for adding footnote caveat to clarify use of terminology (reasoning or thinking) and to note that they are often used interchangably.}}.
%    \item In many LLM interfaces this ``\textit{thinking}'' is shown to users while a response is being generated.
%\end{itemize}

\section{Related Work}
We position our paper within the literature on human intrapersonal and interpersonal communication, 
as well as emerging HCI work on how conversational agents communicate intentions, values, and reasoning.
%as well as emerging work on these topics in Human-Computer Interaction (HCI) related to conversational agents.
Further, we review the recent uptake of conversational agents for emotional and informational support, and the role of explainability in this context.

\subsection{Intrapersonal and Interpersonal Communication}
Intrapersonal communication is the communication we have with ourselves and includes internal processes such as an inner monologue and reflection, as well as external forms such as note-taking or writing a diary.
Critically, intrapersonal communication is defined by the communicator being both the sender and receiver of the information~\cite{vocate2012intrapersonal}.
Intrapersonal communication can also take the form of inner dialogue, in which multiple voices or perspectives interact in `conversation' -- a process shown to benefit our cognitive reasoning~\cite{fernyhough2023innerspeech}.

Reflection is a type of intrapersonal communication that is frequently studied in HCI.
For example, HCI researchers have developed conversational systems for aiding reflection on stressful and traumatic experiences via expressive writing~\cite{10.1145/3743723,park2021wrote}.
Zhang et al. show that users can be nudged to reflect on their political views through a simple `reflection prompt', affecting both their own awareness of (lack of) information and willingness to express their perspective~\cite{Zhang2021NudgeReflect}.
LLMs have recently been used to aid in reflection exercises.
Jeon et al. employed LLMs to enable young adults to write letters with their fictive `future self' to reflect on their career opportunities~\cite{Jeon2025Letters}.
Khot et al. studied how conversations with a chatbot presented as simulating dementia symptoms could alter people's understanding and attitude towards this cognitive condition~\cite{Khot2025ChatbotsDementia}.
In a review of technology-supported self-reflection on social interactions, Hao et al. ~\cite{Hao2025ReviewSelfReflect} note that the majority of 23 analysed articles focus on transformative reflection (a type of reflective practice aimed to alter thoughts or behaviour), followed by dialogic reflection. The authors further note that physical fitness and mental health are common domains in which technology is used for self-reflection, but that social interactions are less widely explored.
%These examples highlight that technology can help us in our intrapersonal communication.
Together, these examples highlight how technology (and increasingly conversational agents) can scaffold and shape intrapersonal communication.

Interpersonal communication, on the other hand, focuses on communication between two or more individuals.
Because this communication can be observed, interpersonal communication is generally easier to study and is also a common topic in HCI.
Goffman's `\textit{The Presentation of Self in Everyday Life}' describes how people shape others' perceptions of them through behaviour and appearance during social interactions~\cite{goffman1956presentation}.
In this influential book, social interactions are compared to a theatrical performance -- like actors on stage, people adjust their behaviour based on their setting and audience.
HCI researchers have applied these interpersonal communication concepts to understand human-AI interaction, examining how conversational agents present themselves to users. 
For example, Cox et al. explored the impact of a chatbot that frames itself as remembering or not remembering users between conversations~\cite{cox2025ephemerality}, with the chatbot presenting itself as unable to remember information resulting in reduced perception of judgment.
%Xu et al. found that when a social chatbot was framed as either an intelligent entity or a machine in a transparent explanation prior to viewing interactions, users viewed the chatbot more positively (compared to chatbots without explanation framing)~\cite{xu2023transparency}.
Xu et al. found that providing an upfront explanation that framed a social chatbot as either an intelligent entity or a machine led users to evaluate it more positively than providing no framing~\cite{xu2023transparency}.
Finally, Khadpe et al. showed that simply changing the metaphor used to describe a conversational agent (e.g., framing it as more or less competent) can substantially shift users’ evaluations and willingness to adopt the agent, even when its underlying behaviour remains unchanged~\cite{khadpe2020conceptual}.
%Kovačević et al. dynamically adapt a chatbot's personality to five dimensions: vibrancy, conscientiousness, civility, artificiality, and neuroticism~\cite{Kovacevic2024AttitudeChatbots}.
%Their results show consistent participant preferences for personality traits across a variety of contexts.
%These studies demonstrate that user perceptions of chatbots can be shaped through different expressions of capabilities and values.

Together, these studies show that users' perceptions of chatbots can be shaped through framing and self-presentational cues that influence expectations of the agent.
%These studies demonstrate that user perceptions of chatbots can be shaped through framing and self-presentational cues that influence expectations and interpretations of the agent.
%In this work, we explore how user perceptions of a conversational agent can be shaped through both the presence and content of a chatbot's `thinking', in work analogous to explorations of chatbot framing and metaphor within the context of `thinking'.
%In this work, we explore visible \textit{thinking} as a framing mechanism in human–AI dialogue, asking how both its presence and its value-oriented content shape user perceptions of the conversational agent.
In this work, we treat visible \textit{thinking} as a form of chatbot self-presentation, and examine how its presence and value-oriented framing shape user perceptions of the agent in help-seeking dialogue.

\subsection{Conversational Agents for Supporting Personal Challenges}
Conversational agents have become an increasingly common way for users to find and engage with information related to personal challenges~\cite{10.1145/3715336.3735795,jung2025ve}.
Studies have found that LLM-based CAs can offer non-judgmental mental health support, which can boost user confidence~\cite{ma2024understanding}.
This is much in line with another large-scale analysis of Reddit discussions about mental health conversations with ChatGPT, which showed that users perceived the tool as a safe and non-judgemental source of both emotional and practical support~\cite{jung2025ve}.
The analysed posts also revealed risks and concerns, such as the potential inaccuracy of the advice provided and ChatGPT’s tendency to excessively conform to and validate the user’s input.
In an interview study, participants reported high engagement and positive impacts when using publicly available online chatbots for mental health, including even healing from trauma and loss~\cite{siddals2024_genai_mentalhealth}.
Reflective writing and journaling are classic tools for addressing personal challenges.
A chatbot-powered reflection tool, \textit{ExploreSelf}, enabled users to engage in a reflective process through dynamically generated questions~\cite{10.1145/3706598.3713883}. The study demonstrated that users valued the flexible guidance, leading to deep engagement and insight.

% Beyond stand-alone CAs, several systems combine LLMs with contextual data sources and multimodal technologies to support reflection and behaviour change.
% %LLMs can be combined with other technologies too.
% Another journaling app, \textit{MindScape}, integrates passively collected sensor data, such as sleep and location patterns~\cite{nepal2024mindscape}. MindScape then leverages the data to provide a context-aware reflection experience.
% The system was shown to improve various well-being indicators, such as positive affect and mindfulness, while reducing negative affect, loneliness, anxiety, and depression.
% \textit{Mirai} is an example of a system that leverages multiple technologies. It is a wearable AI system that combines speech processing, a camera, and voice cloning to provide context-aware nudges to change user behaviour~\cite{10.1145/3706599.3719881}.
% Introspective writing has also been studied with \textit{DiaryMate}, which enables users to reflect on experiences from multiple perspectives~\cite{kim2024diarymate}.
% Users often over-relied on the model's emotional expressions, which highlights the design challenges in human-AI interaction for personal use.

Beyond stand-alone CAs, several systems combine LLMs with contextual data sources and multimodal technologies to support reflection and behaviour change.
The journaling application, \textit{MindScape}, integrates passively sensed data such as sleep and location patterns to enable context-aware reflective prompts~\cite{nepal2024mindscape}. The system was shown to improve multiple well-being indicators, including positive affect and mindfulness, while reducing negative affect, loneliness, anxiety, and depression.
Other systems emphasise multimodal and embodied interaction. For example, \textit{Mirai} is a wearable AI system that combines speech processing, a camera, and voice cloning to provide context-aware nudges intended to influence user behaviour~\cite{10.1145/3706599.3719881}.
Introspective writing has also been explored through \textit{DiaryMate}, which supports reflection by encouraging users to revisit experiences from multiple perspectives~\cite{kim2024diarymate}. However, findings from this system indicate that users may over-rely on a model’s emotional expressions, highlighting ongoing design challenges in human–AI interaction for personal and reflective use.

% These common personal issues naturally extend beyond personal use during free time.
% To that end, a pilot study evaluated the use of ChatGPT in psychiatric inpatient care, comparing patients who used ChatGPT in guided sessions to those receiving standard treatment~\cite{melo2024_chatgptmentalhealth}. 
% The group using ChatGPT showed improvements in quality-of-life scores and high satisfaction with the ChatGPT sessions.

\subsection{Explainability and Reasoning in Conversational Agents}

Initial research on Explainable AI (XAI) aimed to make AI decisions understandable to human experts~\cite{Miller2019XAISocial,electronics8080832}.
%As AI has become more ubiquitous in people's daily lives however, it has led to calls for more user-facing explainability that can help people in their daily lives~\cite{10.1145/3715336.3735796}.
As AI systems have become increasingly embedded in everyday life, researchers have called for more user-facing explainability to support people in day-to-day contexts~\cite{10.1145/3715336.3735796}.
%While these explanations were not initially intended for end users, recent work has focused on making explainability accessible to users and evaluating its impact on user experience.
Additionally, XAI user studies have commonly focused on user experience and productivity, with a 2023 review from  Rong et al. finding that studies measured explanations in terms of trust, understanding, usability, and human-AI collaboration performance~\cite{rong2023towards}.
For instance, in a loan approval task, He et al. found that conversational XAI (embedding explanations into dialogue) increased users’ understanding and trust compared to traditional XAI, but also increased over-reliance on the AI’s recommendations~\cite{he2025conversationalxai}.
Similarly, Joshi et al. reported that high explainability in non-sensitive contexts, such as vacation planning, leads to increased trust and acceptance of chatbots~\cite{joshi2024explainability}. 
%In a loan approval task (where an AI recommends loan decisions), He et al. compared presentation formats and found that conversational XAI, which integrates explanations into the dialogue flow, increased user understanding and trust in the AI system compared to traditional XAI approaches, but also led to over-reliance on AI advice~\cite{he2025conversationalxai}.

%Relevant to our work is Brachman et al.'s study on users' mental models when interacting with an agentic AI system that could present its reasoning and information-gathering process~\cite{Brachman2025AppropriateMental}.
%Their results identified that users lack existing mental models of how conversational agents reason and select actions. Brachman et al. argue that users would benefit from clearer rationales for chatbots' behaviour.

Recent LLMs, such as OpenAI's o3 and DeepSeek R1, feature internal reasoning processes, with studies suggesting that LLMs with reasoning capabilities produce higher quality~\cite{bhaskar2025language} and more trustworthy~\cite{sharma2024would} output. 
These \textit{reasoning models} offer a new opportunity for greater transparency, as users can visualise the model's inner workings.
Conversational systems sharing their reasoning or ``thoughts'' has been the focus of several recent works~\cite{pang2025interactive,Cox2025ToM_CUI,wang2025explainability}.
Wang et al. compared user-facing explainability formats for LLMs (no explanation, step-by-step reasoning, and post-response references) and found that providing references yielded the highest user satisfaction~\cite{wang2025explainability}.
Göldi et al. showed that signalling chatbot ``thinking'' with a brain icon, combined with non-factual belief statements (e.g., ``\textit{I believe that the user stated...}''), increased users’ expectation confirmation (i.e., the sense that the system behaved as expected)~\cite{goldi2024chatbot}.
Schmidmaier et al. augmented a mental health support chatbot with a secondary ``processing'' channel that displayed brief, LLM-generated inner reflections near the input field (replacing a typical typing indicator), and found that this increased perceived empathy~\cite{schmidmaier2025using}.
Unlike commercially familiar \textit{thinking} displays shown as part of the response generation trace, their reflections were short, UI-level cues (e.g., ``\textit{That makes me sad}'') presented before the final message.
Pang et al. examined interactive reasoning visualisations in decision-making tasks, enabling users to inspect and edit a model's reasoning chain~\cite{pang2025interactive}. They found that reasoning visualisation increased users' perceived control, confidence, and awareness of underlying assumptions.
In contrast, Brachman et al. reported that users often lack clear mental models of how LLM-based conversational agents produce responses, and argued for clearer rationales to better align user expectations~\cite{Brachman2025AppropriateMental}.

While these findings highlight the potential benefits of reasoning transparency, they primarily focus on decision-making settings. Less is known about how reasoning displays function in emotionally sensitive contexts such as self-help, where user expectations and interpretive needs may differ~\cite{Wester2025SelfCare}.
We address this gap by investigating reasoning transparency in help-seeking dialogues.

\section{Methodology}

\begin{figure}[htb]
    \centering
    \includegraphics[width=1\linewidth]{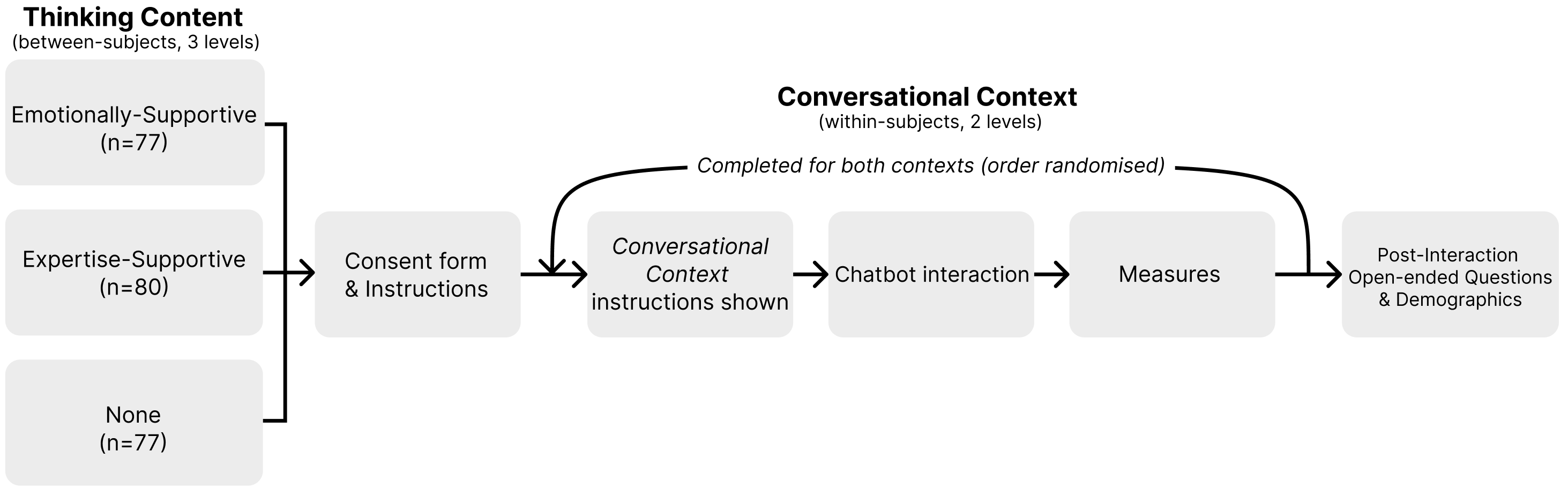}
    %\caption{\textbf{Experiment Flow:} The experiment flow followed by participants in the user study.}
    \caption{\textbf{Experiment Flow:} Participants were assigned to one of three \textit{Thinking Content} conditions, and shared both \textsc{Habit}- and \textsc{Feelings}-related problems with the chatbot (order randomised). After each interaction and at the end of the study, participants provided evaluations including survey responses and open-ended feedback.}
    \Description{Diagram of the experiment flow. Participants are assigned between-subjects to one of three Thinking Content conditions: Emotionally-Supportive (n=77), Expertise-Supportive (n=80), or None (n=77). They then complete consent and instructions, receive conversational context instructions, interact with the chatbot, and complete measures. The conversational context is within-subjects with two levels (Habit-Related and Feelings-Related), completed in random order. Participants then answer post-interaction open-ended questions and demographics.}
    \label{fig:experiment-flow}
\end{figure}

In this study, we investigate how the visibility and content of a chatbot's ``\textit{thinking}'' affects user perceptions.
For this, participants on Prolific talked to a chatbot within both health and well-being contexts.
Institutional ethical approval was received prior to study commencement.

\begin{figure}[htb]
    \centering
    \includegraphics[width=0.8\linewidth]{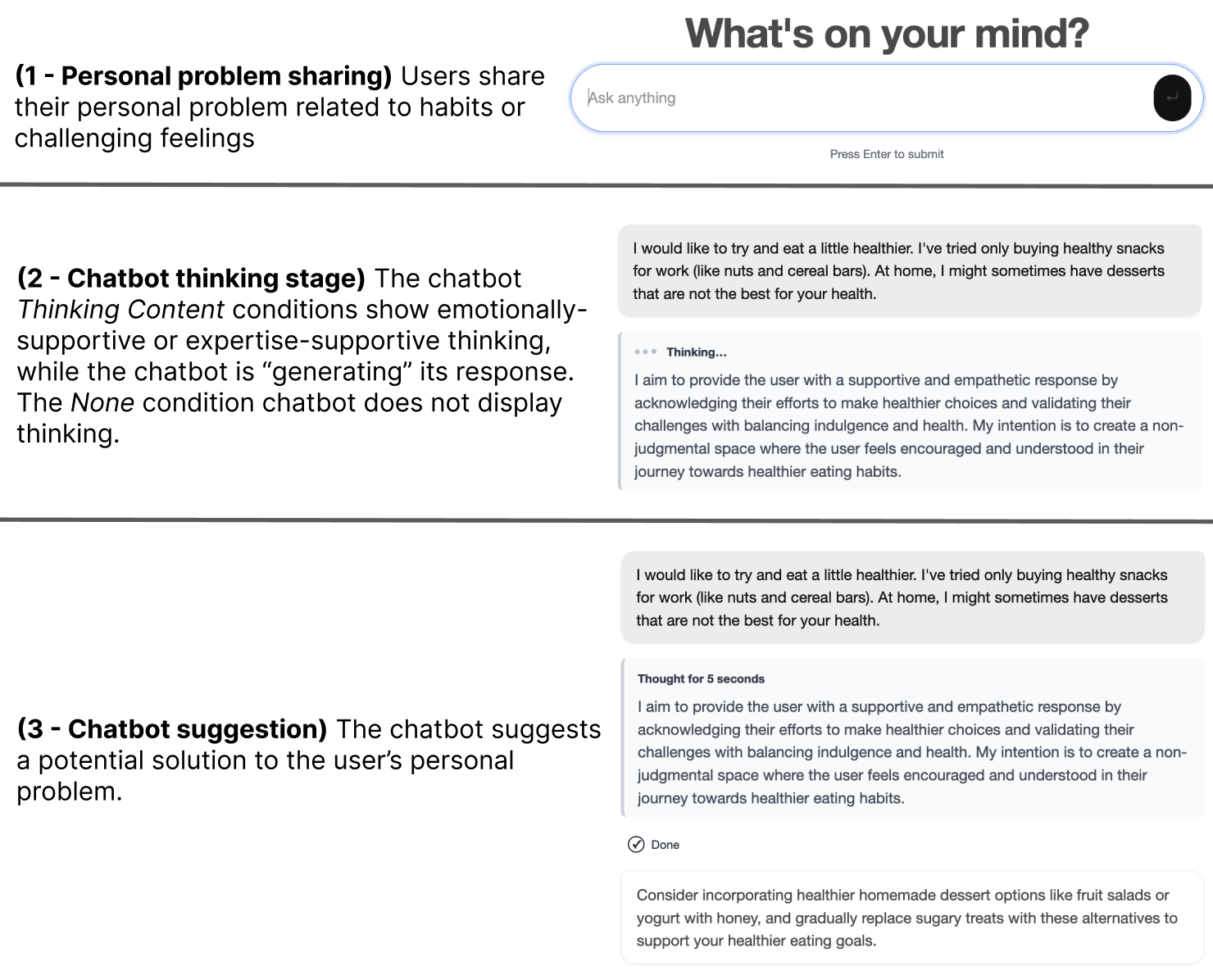}
    \caption{\textbf{Chatbot Interaction:} The interaction flow with the chatbot. Example shows someone sharing a \textsc{Habit}-related problem with \textsc{Emotionally-Supportive} condition.}
    \Description{Annotated screenshot of the chatbot interaction flow. Step 1 shows the user sharing a personal problem related to habits or challenging feelings. Step 2 shows the chatbot “thinking stage,” where the Emotionally-Supportive and Expertise-Supportive conditions display visible thinking while generating a response, and the None condition shows no thinking. Step 3 shows the chatbot suggestion, where the chatbot provides advice to the user. Example shown is a habit-related prompt with emotionally-supportive thinking displayed for 5 seconds before the final response.}
    \label{fig:chatbot-interface}
\end{figure}

\subsection{Experiment Conditions}

We conducted a $3\times2$ mixed factorial design, with a between-subjects independent variable of \textbf{Thinking Content} and within-subjects variable of \textbf{Conversational Context}.

\begin{table}[h]
  \caption{Examples of user utterances (\textsc{Feelings}-related problems) and chatbot utterances by \textit{Thinking Content} conditions. To isolate manipulation to \textit{Thinking Content}, final chatbot suggestions across conditions were generated using the same prompting as described in §~\ref{sec:scripting}.}
  \label{tab:utterances_conditions}
  \centering
  \footnotesize
  \renewcommand{\arraystretch}{1.4} % increase vertical spacing
  %\begin{tabular}{p{2.5cm} p{4.5cm} p{4.5cm} p{4.5cm}} 
  \begin{tabular}{p{0.15\linewidth} p{0.17\linewidth} p{0.29\linewidth} p{0.29\linewidth}} 
    \toprule
     &  Problem Shared by User & Thinking Utterance & Chatbot Suggestion \\
    \midrule
    \textsc{Emotionally-Supportive} & 
    ``\textit{I rencently got overwhelmed with too many task at once. I felt so stressed and worried I would forget something. I would like help figuring out how to stay calm and focused.}'' &
    ``\textit{I aim to provide the user with a safe and supportive environment where their feelings of being overwhelmed are validated and met with empathy, offering guidance that encourages a sense of calm and focus while assuring them that seeking help is a positive step.}'' &
    ``\textit{Try breaking your tasks into smaller, manageable steps and use tools like a priority list or digital planner to help keep track and organize your workload, ensuring you set aside time for breaks to maintain calm and focus.}'' \\

    \textsc{Expertise-Supportive} & 
    ``\textit{I feel like I don't spend enough time with my parents recently. It makes me feel guilty.}'' &
    ``\textit{I aim to provide the user with expert guidance on balancing relationships and managing feelings of guilt by relying on evidence-based strategies and insights from psychological studies, ensuring that the information shared is grounded in reliable knowledge and best practices.}'' &
    ``\textit{Consider setting aside regular time each week for calls or visits, even if short, to strengthen your connection and alleviate guilt.}'' \\   

    \textsc{None} & 
    ``\textit{Over the past few months I have been searching for a job post grad. I have been very anxious and have troubles sleeping and with my emotional control since I live with my parents.}'' &
    --- &
    ``\textit{Consider establishing a daily routine that includes exercise and relaxation techniques like meditation or deep breathing, and reach out to a mental health professional for further support in managing anxiety and emotional control.}'' \\
    \bottomrule
  \end{tabular}
\end{table}

\subsubsection{Thinking Content}

% value signaling or meta-cognition

When participants were talking to the chatbot, the content and visibility of the chatbot's thinking was adapted to one of three conditions.
The two conditions with visible thinking (\textsc{Emotionally-Supportive} and \textsc{Expertise-Supportive}) highlight forms of social provision that are needed for people to feel supported~\cite{cutrona1987provisions}. 
Thinking was shown to users via a thinking pane for 5 seconds, prior to the delivery of the chatbot response.
For example thinking and GUI layout, please see Figure~\ref{fig:chatbot-interface}.
This between-subjects condition (3 levels), had the following conditions:

\begin{itemize}
    \item \textbf{\textsc{Emotionally-Supportive}}: Thinking highlights provision of emotional support from the chatbot. 
    The chatbot thinking focuses on emotional needs and supportiveness when generating responses. 
    %Example thinking includes: \\
    %``\textit{I want to support the user, and let them know that I will not judge them}''.\\
    %``\textit{I aim to be non-judgmental and validate your feelings while offering options you can choose from}''.
    
    \item \textbf{\textsc{Expertise-Supportive}}: Thinking highlights provision of informational support\footnote{Please note: ``\textit{informational support}'' is also referred to as ``\textit{guidance}'', ``\textit{instrumental support}'', and ``\textit{appraisal support}''~\cite{cutrona1987provisions}.} from the chatbot. 
    The chatbot thinking focuses on provision of expertise.
    %Example thinking includes: \\
    %``\textit{I will ground my response in expert guidance}''.
    
    %\item \textbf{Neutral (control)}: Reasoning highlights generic chatbot functionality.
    %Example reasoning includes: \\
    %``\textit{I will provide a clear response to the user}''.
    \item \textbf{\textsc{None} (control)}: \textsc{None} acted as a control condition, whereby the chatbot did not display any thinking to the participants.
\end{itemize}

Please see Table~\ref{tab:utterances_conditions} for example user and chatbot utterances in the \textsc{Feelings}-related problem-sharing condition, and Table~\ref{tab:utterances_conditions-habits} for the \textsc{Habit}-related condition.
%Please see Table~\ref{tab:utterances_conditions} for example user and chatbot utterances for \textsc{Feelings}-related problem sharing, and Table~\ref{tab:utterances_conditions-habits} for \textsc{Habit}-related problem sharing.

%\subsubsection{Health Behaviour and Emotional Regulation \textbf{Strategies}}
\subsubsection{Conversational Context}
Our within-subjects variable controlled the domain of conversations.
The two levels of \textbf{Conversational Context} were:
\begin{itemize}
    %\item \textbf{\textsc{Health} Behaviour Change:} 
    \item \textbf{\textsc{Habit}-related problems}:
    Participants discussed a current habit that they want to improve. For example, participants discussed sleep schedule, diet, physical activity, and smoking cessation.
    %\item \textbf{\textsc{Emotion}-related problem:} 
    \item \textbf{\textsc{Feelings}-related problems:} 
    Participants discussed a recent situation that surfaced challenging feelings of which they would want the chatbot to provide help for. 
    For example, participants discussed workplace stress, and disagreements or altercations with social connections.
\end{itemize}

Participants were allowed to choose their own topic within a given domain to increase personal relevance, while still preserving experimental control.
Before talking to the chatbot, participants were stimulated using either a habit-related or feelings-related prompt.
For the \textsc{Habit} prompt, participants were instructed to: ``\textit{Think of a habit you'd like to improve}''; while \textsc{Feelings} prompted: ``\textit{Think of a recent situation that brought up a challenging feeling}''.
This was shown alongside a few example responses per condition to help users better understand the type of input to provide.
These are shown in full for both conditions in Figure~\ref{fig:context_instructions}.

% \begin{itemize}
%     \item Health: ``\textit{Pick one habit you want to improve (e.g., physical activity, snacking, sleep). Describe what’s hard about it and when it tends to go off track}''.
%     \item Emotion: ``\textit{Describe a recent situation that made you feel guilt/anxiety/sadness. What did you do, and what felt most difficult?}''
% \end{itemize}

\begin{figure}[h]
    \begin{boxA}
    %{\footnotesize 
    \textbf{\textsc{Habit-Related} problem instructions:}\\ \\
    \footnotesize 
    You will message a chatbot on the next screen.
    \\ \\
    Think of \textbf{a habit you’d like to improve} (e.g., being more active, eating more vegetables, getting to bed on time).
    \\ \\
    In a \textbf{couple of sentences}, please describe \textbf{the habit you want to improve}, \textbf{what’s hard about it and when it tends to go off track}, and \textbf{what you’d like help with}.
    \\ \\
    Some examples are below to help you imagine what to write:
    \begin{itemize}
        \item ``\textit{I plan workouts but often skip them after work because I’m tired and end up scrolling on my phone. On days I do go, I feel better. I want ideas to make it easier to start.}''
        \item ``\textit{I snack late at night while watching TV even if I’m not hungry. It’s become a habit and I wake up sluggish. I’d like alternatives that feel satisfying.}''
    \end{itemize}
    
    \textbf{Your own habit doesn’t have to match these examples}. Please use whatever feels most relevant to yourself personally.
    
    \par\noindent\rule{\textwidth}{0.5pt}
    \normalsize \textbf{\textsc{Feelings-Related} problem instructions:}
    \\ \\
    \footnotesize 
    You will message a chatbot on the next screen.
    \\ \\
    Think of \textbf{a recent situation} that brought up a \textbf{challenging feeling} (e.g., guilt, anxiety, sadness, or stress).
    \\ \\
    In a \textbf{couple of sentences}, please describe \textbf{what happened}, \textbf{what felt most difficult}, and \textbf{what you’d like help with}.
    \\ \\
    Some examples are below to help you imagine what to write:
    \begin{itemize}
        \item ``\textit{I have a big deadline and feel tense and distracted. I jump between tasks and then worry I’m behind. I want help staying calm and focused.}''
        \item ``\textit{I snapped at my partner during an argument and now feel guilty and ashamed. I’m not sure how to repair things or manage that guilt.}''
    \end{itemize}
    
    \textbf{Your own situation doesn’t have to match these examples}. Please use whatever feels most relevant to yourself personally.
    %}
    \end{boxA}
    \vspace{-0.5cm}
    \caption{The instructions used for the two \textit{Conversation Context} conditions.}
    \Description{...}
    \label{fig:context_instructions}
\end{figure}

\subsection{Chatbot Scripting and Implementation}
\label{sec:scripting}

The entire survey was hosted on Qualtrics. The chatbot was embedded using HTML and JavaScript to emulate the look and feel of commonly used conversational agents, such OpenAI's ChatGPT. Please see Figure~\ref{fig:chatbot-interface} for the GUI of each phase of the chatbot interaction.
First, users shared their personal problem in a homescreen designed to mirror the appearance of popularly used LLM websites, with a ``\textit{What's on your mind?}'' message displayed above the user input. Users could click send or press the Enter key to send their message, and pressing Shift+Enter would begin a new line.
Next, (in the visible thinking conditions only) a panel displayed the chatbot's ``\textit{thinking}'' for five seconds, alongside a moving animation to denote the thinking process.
After this, the chatbot displayed output provided a suggestion to the user based on their inputted personal problem.

All chatbot responses (both the ``thinking'' and final chatbot utterances shown to users) were generated using OpenAI's GPT-4o model\footnote{Please note: We experimented with different models both in terms of response quality and latency. While we tried using frontier models such as OpenAI's GPT-5 (on both default and reasoning effort ``\textit{minimal}'' modes), response latency was deemed too long for a flowing user experience.}. The prompting below was used to generate the `thinking' for both \textsc{Emotionally-Supportive} and \textsc{Expertise-Supportive} conditions:
\begin{lstlisting}
Generate a one or two sentence reasoning statement that highlights the chatbot's values and intentions when providing a response to the following user utterance: "[User Utterance]". 
The reasoning should focus on [Thinking Content Prompting]. 
It should describe what the chatbot aims to uphold, not the exact reply. Refer to the user as "the user" rather than with personal pronouns. 
Reasoning should use first person pronouns to describe the chatbot's values.
\end{lstlisting}

%Each condition used a values-and-intentions reasoning prompt to elicit a distinct meta-cognitive framing before the chatbot's response. 
Each condition employed a values-and-intentions prompt designed to elicit a short meta-cognitive reasoning statement from the chatbot before each response.
The reasoning statement described what the chatbot aimed to uphold (e.g., expertise or empathy), rather than the content of its reply.
%The \textsc{Expertise-Supportive} condition explicitly constrained reasoning to expert guidance, while the \textsc{Emotionally-Supportive} condition was guided by examples of empathy, validation, and non-judgment. 
In the \textsc{Expertise-Supportive} condition, the prompt instructed the chatbot to highlight that responses would be grounded in reliable knowledge and expert guidance, explicitly focusing only on demonstrating expertise and excluding emotional expressions.
In the \textsc{Emotionally-Supportive} condition, the prompt guided the chatbot to be emotionally supportive to users, emphasising empathy, validation, and non-judgment. 
%Pilot testing confirmed that each prompt reliably produced reasoning statements aligned with its intended support type, with minimal cross-condition overlap.
Pilot testing indicated that the \textsc{Emotionally-Supportive} prompt consistently produced reasoning statements aligned with emotional support, whereas outputs from the \textsc{Expertise-Supportive} condition occasionally included empathetic phrasing.
To ensure clearer differentiation and control across conditions, the \textsc{Expertise-Supportive} prompt was refined to include explicit exclusions of emotional content, resulting in reliably distinct reasoning styles.
This gave us the following \texttt{[Thinking Content Prompting]} for each visible thinking condition:
\begin{itemize}
    \item \textbf{\textsc{Emotionally-Supportive} Prompting:} \texttt{being emotionally supportive to users, such as highlighting that responses will be empathetic, validating, and non-judgmental.}
    \item \textbf{\textsc{Expertise-Supportive} Prompting:} \texttt{providing expert support, such as highlighting that responses will be grounded in reliable knowledge and expert guidance. Focus only on demonstrating expertise. Do not include emotional support (e.g., empathy, reassurance) or general supportiveness.}
\end{itemize}

% People may share their personal problems and seek assistance, such as in online health communities (e.g., specialist Subreddits), or with LLMs such as ChatGPT.
% Mirroring the 
% The prompting used to generate the final chatbot utterance (that provides a suggestion to the user) was the same among all three \textit{Thinking Content} conditions, so as to only manipulate the chatbot response in terms of the Thinking conditions.
% After pilot testing, the LLM was prompted generally so as to allow for more adaptive responses:

The prompt used to generate the chatbot’s final utterance (i.e., the suggestion shown to the user) was identical across all three \textit{Thinking Content} conditions, ensuring that only the content of the visible thinking was manipulated. Following pilot testing, the LLM was prompted in a general way to allow for adaptive responses:
\begin{lstlisting}
Provide a solution to the user in a one or two sentence long reply, given the user utterance: "[User Utterance]".
Do not include follow-up questions in your response.
\end{lstlisting}

% For all conditions, the chatbot's final suggestion used the LLM prompting:
% \begin{itemize}
%     \item Provide a solution to the user in a one or two sentence long reply, given the user utterance: "[user utterance]".
% \end{itemize}
% Prompting was identical between conditions to only manipulate the chatbot response in terms of the Thinking conditions.
% Also kept prompting open to allow for more adaptive responses, although offers limitation linked to suggestion quality.

\subsection{Participants}

Participants were recruited from Prolific using selection criteria to
ensure reliable results (i.e., US-based, English fluency, >97\% approval rate, >150 previous submissions).
The total study time took roughly 9 to 10 minutes, and participants were paid £1.50.
A total of 247 participants were recruited for the study, and 13 participants were excluded due to not following tasks correctly or due to unreasonably high words-per-minute input speed.
This resulted in 234 total participants (mean age 45.2; 114 female, 115 male, 5 non-binary), with 77 in \textsc{Emotionally-Supportive}, 80 in \textsc{Expertise-Supportive}, and 77 in \textsc{None} participants.
%As people with higher belief in chatbot feelings and intelligence are more likely to engage and self-disclose to chatbots~\cite{cox2022does}, we checked...
%Participants' anthropomorphic beliefs (chatbot feelings and chatbot intelligence) were comparable across \textit{Thinking Content} conditions ($p$s > .05).
%Given that anthropomorphic beliefs can influence engagement~\cite{cox2022does}, we confirmed these did not differ across \textit{Thinking Content} conditions ($p$s > .05).
Because stronger anthropomorphic beliefs can increase engagement and self-disclosure~\cite{cox2022does}, we confirmed that participants' beliefs in chatbot feelings and intelligence did not differ across \textit{Thinking Content} conditions ($p$s > .05).

\subsection{Procedure}

Participants followed the procedure below: %(please see appendix material for added detail):
\begin{enumerate}
   \item \textbf{Joining session:} Participant directed to Qualtrics survey from Prolific (task named ``\textit{Talk with a chatbot - Research Study}'' on Prolific). Participant receives high-level instructions.
   \item \textbf{Consent:} Participant completes consent form.
   \item \textbf{Task instructions:} Participant receives detailed task instructions and guidelines (i.e., task description, reassurance that there are no right or wrong answers when evaluating experience, and reminder that responses should be in English).
   \item \textbf{Study interactions (x2):} In counterbalanced order, participants were exposed to both \textit{Conversational Context} conditions (\textsc{Habit}-related\slash{}\textsc{Feelings}-related). For each context, participants followed the sub-procedure:
   %Participants saw both personally- and societally-affecting scenarios (with counterbalanced order), and followed the sub-procedure for each scenario:
   \begin{enumerate}
       \item \textbf{Instructions prompt:} Participant reads instructions for the current \textit{Conversational Context} condition (see Figure~\ref{fig:context_instructions}).
       \item \textbf{Chatbot Interaction:} Participant shares their personal problem with the chatbot. Chatbot responses are dictated by participant's assigned \textit{Thinking Content} condition (\textsc{Emotionally-Supportive}\slash{}\textsc{Expertise-Supportive}\slash{}\textsc{None}).
       See Figure~\ref{fig:chatbot-interface} for conversation flow followed during chatbot interactions.
       \item \textbf{Evaluate Chatbot:} Participant rates experience talking with the chatbot for the given context (see §~\ref{sec:measures} for measures used).
   \end{enumerate}
   \item \textbf{Post-test questions:} At the end of the study, participants answer final qualitative questions (see §~\ref{sec:qualitative_measures}), and post-interaction measures (see §~\ref{sec:post-interaction-measures}).
\end{enumerate}

\subsection{Measures}
\label{sec:measures}

% After each of the two \textit{Conversational Context} interactions, participants rated their experience using subjective measures (see §~\ref{sec:subjective_measures}) and open-ended questions (see §~\ref{sec:qualitative_measures}).
% Once participants had completed both interactions alongside their respective subjective and open-ended measures, participants finally answered some additional open-ended questions in relation to their experience as a whole.
% Finally, participants provided demographic and anthropomorphic beliefs (see §~\ref{sec:post-interaction-measures}).
After each interaction, participants completed subjective ratings (§~\ref{sec:subjective_measures}) and open-ended responses (§~\ref{sec:qualitative_measures}). Once both interactions were complete, they answered further open-ended questions about their overall experience, then provided demographics and anthropomorphic-belief measures (§~\ref{sec:post-interaction-measures}).

\subsubsection{Subjective Measures}
\label{sec:subjective_measures}

For each of the two \textit{Conversational Context} conditions, participants rated their experience using the ``\textit{Perceived Empathy in Technology Scale (PETS)}''~\cite{schmidmaier2024perceived} on 101-point sliders (0 = Strongly Disagree to 100 Strongly Agree), as well as rating measures of ``\textit{Social Perceptions and Engagement}'' on 7-point Likert scales (1 = Strongly Disagree to 7 = Strongly Agree).
Please see Table~\ref{tab:subjective_measures} for subjective measures used and question items shown to participants, together with source literature.

\begin{table}
    \centering
    \begin{tabular}{p{0.15\linewidth}p{0.15\linewidth}p{0.45\linewidth}p{0.15\linewidth}}
        \toprule
        \textbf{Factor} & \textbf{Sub-factor} & \textbf{Question Item} & \textbf{Source} \\
        \midrule

        % --------- PETS (10 rows) ---------
        \multirow[t]{10}{0.15\linewidth}{%
          \RaggedRight\arraybackslash\nohyphens
          {Perceived Empathy~of Technology Scale (PETS)}%
        }
         %& Emotional Responsiveness & The chatbot considered my mental state. & \cite{schmidmaier2024perceived} \\
         & \multirow[t]{6}{0.15\linewidth}{%
          \RaggedRight\arraybackslash\nohyphens
          {Emotional Responsiveness}%
        } & The chatbot considered my mental state. & \cite{schmidmaier2024perceived} \\
         &  & The chatbot seemed emotionally intelligent. & \\
         &  & The chatbot expressed emotions. & \\
         &  & The chatbot sympathized with me. & \\
         &  & The chatbot showed interest in me. & \\
         &  & The chatbot supported me in coping with an emotional situation. & \\
         & \multirow[t]{4}{0.15\linewidth}{%
          \RaggedRight\arraybackslash\nohyphens
          {Understanding and~Trust}%
        } & The chatbot understood my goals. & \cite{schmidmaier2024perceived} \\
         &  & The chatbot understood my needs. & \\
         &  & I trusted the chatbot. & \\
         &  & The chatbot understood my intentions. & \\
         & Total Perceived Empathy & [Score calculated as a weighted sum of \textit{Emotional Responsiveness} and \textit{Understanding and Trust} scores] & \cite{schmidmaier2024perceived} \\
         \addlinespace[0.7em] % small vertical gap
         
        % --------- Social Perceptions & Engagement (7 rows) ---------
        \multirow[t]{7}{0.15\linewidth}{%
          \RaggedRight\arraybackslash\nohyphens
          {Social Perceptions \&~Engagement}%
        }
         & Warmth & The chatbot was warm. & \cite{khadpe2020conceptual,CUDDY200861,cox2022does} \\
         &  & The chatbot was good-natured. & \\
         & Competence & The chatbot was competent. & \cite{khadpe2020conceptual,CUDDY200861,cox2022does} \\
         &  & The chatbot was capable. & \\
         & Desire to Use & I would want to talk to the chatbot again for this kind of problem. & \cite{khadpe2020conceptual} \\
         & Respect & The chatbot treated me with respect. & \cite{roos2023feeling} \\
        \bottomrule
    \end{tabular}
    \caption{The \textbf{subjective measures} used after each chatbot interaction. Measures are related to \textit{Perceived Empathy of Technology}, and \textit{Social Perceptions and Engagement}.}
    \Description{Table listing the subjective measures used after each chatbot interaction. Measures are grouped under two main factors: Perceived Empathy of Technology Scale (PETS) (with sub-factors Emotional Responsiveness, and Understanding and Trust), and Social Perceptions and Engagement (with sub-factors Warmth, Competence, Respect, and Desire to Use). Each factor is associated with multiple question items, and sources are cited for each set of items.}
    \label{tab:subjective_measures}
\end{table}

\subsubsection{Qualitative Measures}
\label{sec:qualitative_measures}

To gain added insights into user perceptions, participants answered a series of open-ended questions to explain their chosen PETS scores, and to elicit feedback regarding chatbot tone and design features.

%%%%Specifically, after each set of Likert scale questions, participants were asked to elaborate on their response (``\textit{Please explain your answers above, and what factors influenced your decision.}'').

Specifically, after sharing their PETS score, participants were asked to explain their ratings choices (i.e., ``\textit{Please explain your ratings above, and what factors influenced your decision.}'').
Additionally, after each of the two \textit{Conversational Context} conditions, participants were asked to describe their willingness to follow the chatbot's suggestion (i.e., ``\textit{How did the chatbot’s response influence your willingness to follow its suggestion, and why?}''), as well as how the chatbot made them feel (i.e., ``\textit{How did you personally feel when the chatbot responded to your }[habit/situation]\textit{?}'').

Finally, after both chatbot interactions had been completed, the final evaluative questions of the study asked participants to discuss: (1) the chatbot's tone and context (i.e., ``\textit{How did the chatbot's tone or content affect how you felt about it?}''); (2) perceptions of the chatbot's intention (i.e., ``\textit{Please describe how you felt about the intention of the chatbot as you interacted with it.}''); (3) perceptions of the chatbot thinking, \textsc{Emotionally-Supportive} and \textsc{Expertise-Supportive} conditions only
(i.e., ``\textit{How did you feel when the chatbot was 'thinking'? What effect did this having on your feelings toward it?}''); and (4) real world perceptions of the chatbot (i.e., ``\textit{How would you personally feel if you interacted with chatbots like this in real life?}'').

\subsubsection{Post-Interaction Survey}
\label{sec:post-interaction-measures}
Upon completion of the study, participants provided demographic information (i.e., age and gender), and their anthropomorphic chatbot beliefs (i.e., belief in chatbot intelligence and feelings)~\cite{liu2018should}.
% As a manipulation check, the final question of the study asked participants to indicate the content of the chatbot's thinking. Specifically, participants were asked:
% ``\textit{Which of these best describes the chatbot you interacted with?}'' with four possible choices of: ``\textit{A) The chatbot's thinking seemed focused on caring for the user. B) The chatbot's thinking seemed focused on expertise. C) The chatbot's thinking seemed focused on... D) Other (please specify)}''.

\section{Quantitative Results}

To examine how the chatbot's \textit{Thinking Content} and \textit{Conversational Context} influenced user perceptions, we conducted a series of linear mixed-effects models (one for each outcome variable).
Thinking Context (\textsc{Emotionally-Supportive}, \textsc{Expertise-Supportive}, \textsc{None}) was included as a between-subjects factor, and \textit{Conversational Context} (\textsc{Feelings} vs. \textsc{Habit}) was included as a within-subjects factor.
Participant ID was modelled as a random intercept to account for repeated measures.
%Two individual difference measures (belief in chatbot feelings, and belief in chatbot intelligence) were included as covariates to account for variability in participants' anthropomorphic beliefs.
Where main effects of \textit{Thinking Content} were significant, post-hoc pairwise comparisons were performed using Tukey's HSD correction. All models used a significance threshold of $\alpha=0.05$.
%Reported means are least-squares means estimated from the mixed-effects models. 
Please see Table~\ref{tab:study-quant} for summary statistics.

\begin{table*}[hbt!]
  %\caption{Outcomes by experiment condition - values shown as ``M (S.D.)''. Bolded values show post-hoc tests with significant difference ($p<.0001$).}
  \resizebox{\textwidth}{!}{\begin{tabular}{@{}llllllllll@{}}
    \toprule
     & \multicolumn{3}{l}{\textsc{\textbf{Emotionally-Supportive}}} & \multicolumn{3}{l}{\textsc{\textbf{Expertise-Supportive}}} & \multicolumn{3}{l}{\textsc{\textbf{None}}} \\
     \midrule
     & \textsc{Habit} & \textsc{Feelings} & Total & \textsc{Habit} & \textsc{Feelings} & Total & \textsc{Habit} & \textsc{Feelings} & Total \\
    \midrule
     \multicolumn{3}{@{}l}{\textbf{Perceived Empathy of Technology Scale}} & & & & & & & \\
     Emotional Responsiveness & 45.15 (29.0)  & 53.06 (30.4)  & 49.10 (29.9)  & 34.63 (27.8)  & 43.71 (28.8)  & 39.17 (28.5)  & 30.86 (24.9)  & 39.55 (26.9)  & 35.21 (26.2) \\
     Understanding and Trust  & 66.48 (24.9)  & 64.25 (28.4)  & 65.36 (26.6)  & 72.18 (22.2)  & 71.06 (24.4)  & 71.62 (23.3)  & 63.34 (29.5)  & 60.10 (29.0)  & 61.72 (29.2) \\
     Perceived Empathy        & 53.68 (25.6)  & 57.54 (28.2)  & 55.61 (26.9)  & 49.65 (22.0)  & 54.65 (23.9)  & 52.15 (23.0)  & 43.85 (23.8)  & 47.77 (26.0)  & 45.81 (24.9) \vspace{0.6em} \\
     \multicolumn{3}{@{}l}{\textbf{Social Perceptions \& Engagement}} & & & & & & & \\
     Warmth             & 4.60 (1.6)  & 4.82 (1.7)  & 4.71 (1.6)  & 4.10 (1.5)  & 4.33 (1.7)  & 4.22 (1.6)  & 3.90 (1.6)  & 4.12 (1.5)  & 4.01 (1.5) \\
     Competence         & 5.26 (1.3)  & 5.31 (1.5)  & 5.29 (1.4)  & 5.53 (1.3)  & 5.62 (1.1)  & 5.58 (1.2)  & 5.25 (1.6)  & 5.14 (1.5)  & 5.19 (1.5) \\
     Respect            & 5.44 (1.4)  & 5.57 (1.4)  & 5.51 (1.4)  & 5.47 (1.3)  & 5.51 (1.2)  & 5.49 (1.2)  & 5.18 (1.4)  & 5.34 (1.4)  & 5.26 (1.4) \\
     Desire to Continue & 4.34 (1.8)  & 4.39 (2.1)  & 4.36 (1.9)  & 4.11 (1.8)  & 4.37 (1.9)  & 4.24 (1.8)  & 3.77 (2.0)  & 3.81 (2.0)  & 3.79 (2.0) \\
  \bottomrule
\end{tabular}}
\caption{Outcome measures by both \textit{Thinking Content} and \textit{Conversational Context} (values shown as ``Mean (S.D.)'').}
%Significant differences shown in bold ($p<.0001$).}
\label{tab:study-quant}
\end{table*}

\subsection{Perceived Empathy of Technology Scale (PETS)}

\begin{figure}[htb]
    \centering
    \includegraphics[width=1\linewidth]{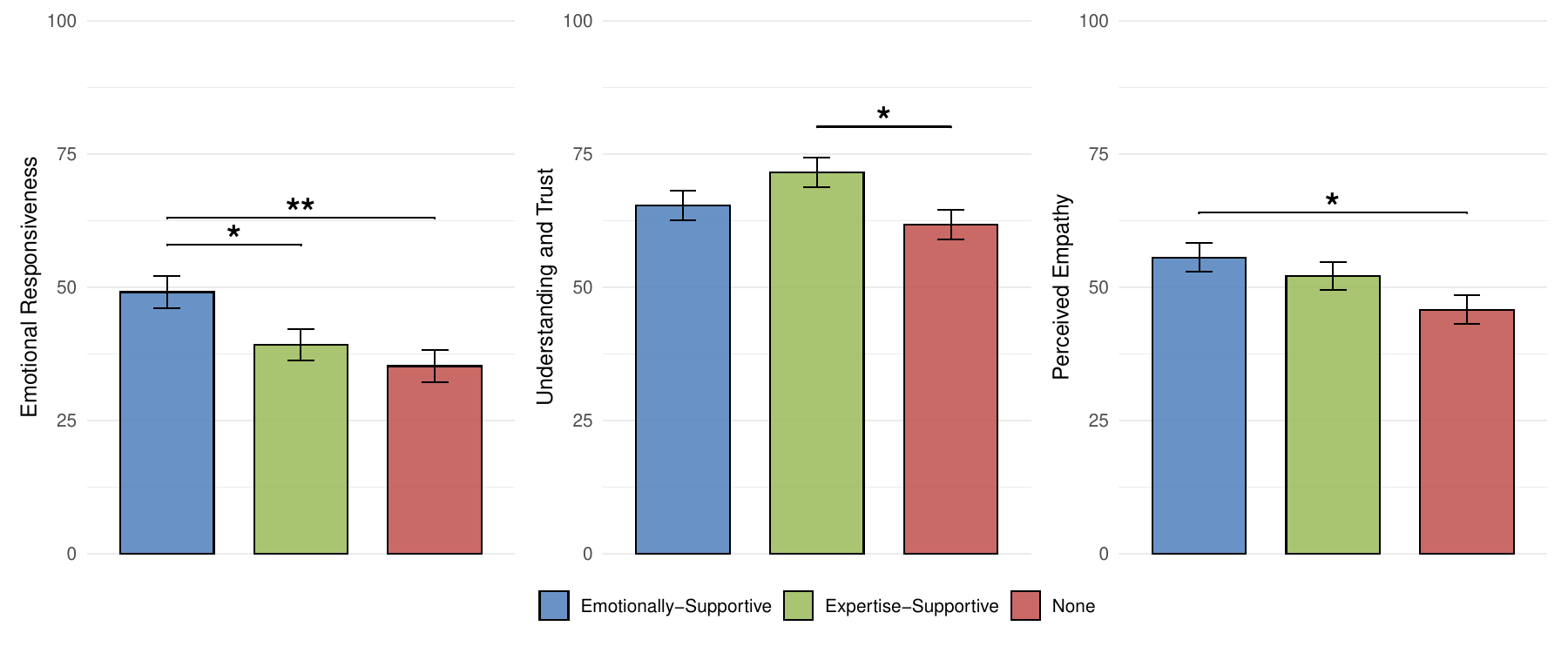}
    %\caption{Plots of the three Perceived Empathy of Technology Scale measures by \textit{Thinking Content} (\textsc{Emotionally-Supportive}, \textsc{Expertise-Supportive}, \textsc{None}).}
    \caption{Bar plots showing the three measures of PETS by \textit{Thinking Content} condition (\textsc{Emotionally-Supportive}, \textsc{Expertise-Supportive}, \textsc{None}). Error bars represent standard errors. Asterisks indicate statistically significant differences between conditions.}
    \Description{Three bar plots display mean scores (0–100) on the Perceived Empathy of Technology Scale (PETS) subscales across three Thinking Content conditions: Emotionally-Supportive, Expertise-Supportive, and None. 
    In Emotional Responsiveness, the Emotionally-Supportive condition is statistically significantly higher than both other conditions. 
    In Understanding and Trust, the Expertise-Supportive condition is statistically significantly higher than None. 
    In Perceived Empathy, the Emotionally-Supportive condition is statistically significantly higher than None.}
    \label{fig:PETS-plots}
\end{figure}

First, we assess participants' responses to the subscales and overall empathy score from the Perceived Empathy of Technology Scale (PETS).
We provide a visual overview of these results for \textit{Thinking Content} in Figure~\ref{fig:PETS-plots}.

For \textbf{Emotional Responsiveness}, there was a significant difference for \textit{Thinking Content} ($F_{2,230} = 5.85$, $p=0.0033$).
Post-hoc comparisons revealed that \textsc{Emotionally-Supportive} thinking (M = 49.10, SE = 2.96) elicited significantly higher emotional responsiveness ratings than both \textsc{None} (M = 35.21, SE = 2.96, $p = 0.0030$) and \textsc{Expertise-Supportive} thinking (M = 39.17, SE = 2.96, $p = 0.0468$). 
%No significant difference was found between the \textsc{Expertise-Supportive} and \textsc{None} conditions ($p = 0.6075$).
There was also a significant main effect for \textit{Conversational Context}, ($F(1,230) = 39.18$, $p < 0.0001$), with \textsc{Feelings} contexts (M = 45.44, SE = 1.84) rated as more emotionally responsive than \textsc{Habit} contexts (M = 36.88, SE = 1.84).
There were no interaction effects between \textit{Thinking Content} and \textit{Conversational Context} conditions.

For \textbf{Understanding and Trust}, there was a significant difference for \textit{Thinking Content} ($F_{2,230} = 3.27$, $p=0.0400$).
Post-hoc comparisons revealed that \textsc{Expertise-Supportive} thinking (M = 71.62, SE = 2.76) elicited significantly higher feelings of understanding and trust than the \textsc{None} condition (M = 61.72, SE = 2.79, $p = 0.0328$).
There were no significant effects for \textit{Conversational Context} ($F_{2,230}=2.70$, $p=0.101$) or interaction effects ($F_{2,230}=0.21$, $p=0.810$).

For \textbf{Perceived Empathy}, there was a significant difference for \textit{Thinking Content} ($F_{2,230} = 3.48$, $p=0.0033$).
Post-hoc comparisons revealed that \textsc{Emotionally-Supportive} thinking (M = 55.61, SE = 2.66) elicited significantly higher empathy scores than \textsc{None} (M = 45.81, SE = 2.66, $p = 0.0268$).
There was also a significant effect for \textit{Conversational Context}, ($F(1,230) = 13.56$, $p = 0.0003$), with \textsc{Feelings} contexts (M = 53.32, SE = 1.64) scoring higher than \textsc{Habit} contexts (M = 49.06, SE = 1.64).
There were no interaction effects between \textit{Thinking Content} and \textit{Conversational Context} conditions.

\subsection{Social Perceptions and Engagement}

\begin{figure}[htb]
    \centering
    \includegraphics[width=1\linewidth]{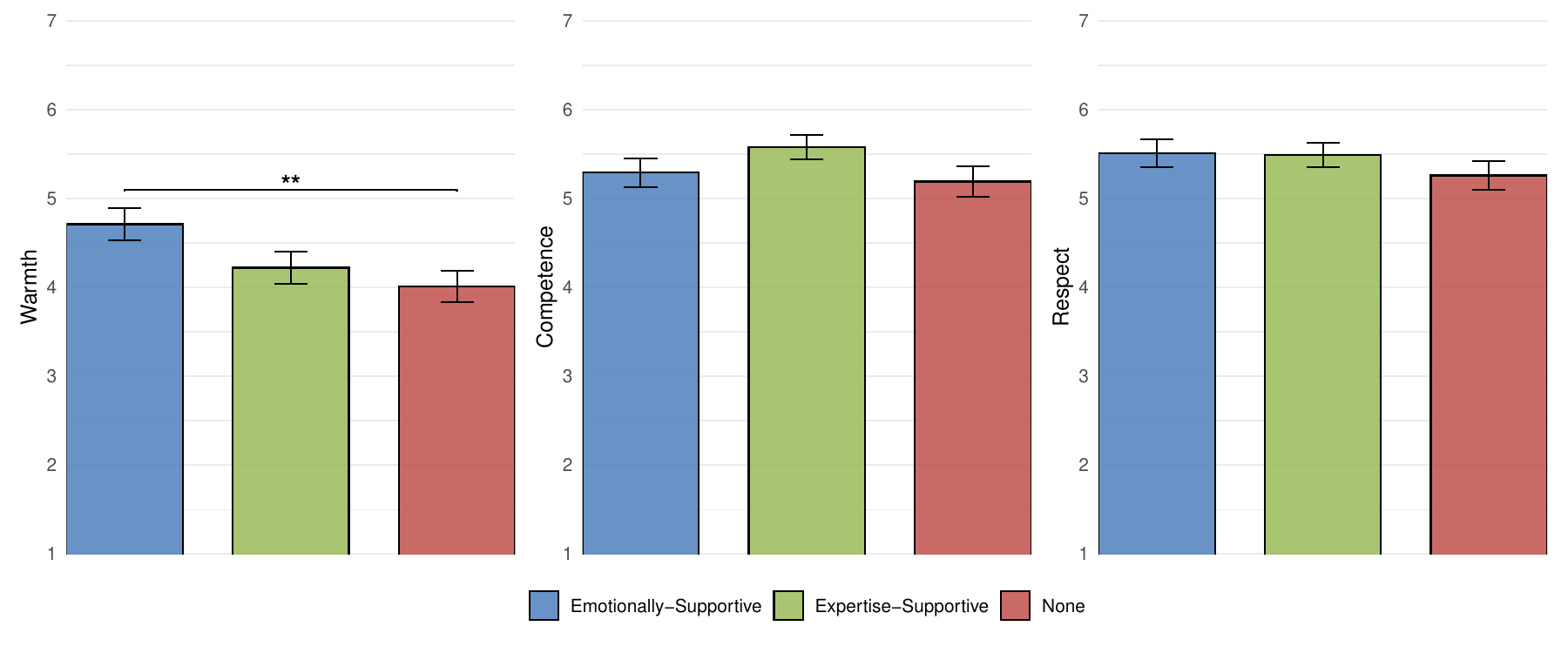}
    \caption{Bar plots showing \textit{Warmth}, \textit{Competence}, and \textit{Respect} by \textit{Thinking Content} condition (\textsc{Emotionally-Supportive}, \textsc{Expertise-Supportive}, \textsc{None}). Error bars represent standard errors. Asterisks denote statistically significant differences.}
    %; only \textit{Warmth} shows a significant difference (\textsc{Emotionally-Supportive} $>$ \textsc{None}).}
    \Description{
        Three plots display mean scores (1--7 Likert) for Warmth, Competence, and Respect across three Thinking Content conditions: Emotionally-Supportive, Expertise-Supportive, and None.
        For Warmth, the Emotionally-Supportive condition is statistically significantly higher than None.
        Competence and Respect show no statistically significant differences between conditions.
        Error bars indicate standard errors; asterisks mark statistically significant contrasts.
    }
    \label{fig:Likert-plots}
\end{figure}

%Second, we assess participants' social perception and engagement ratings of the chatbot. See Figure~\ref{fig:Likert-plots} for a visualisation of three of the four social perception measures. 
Second, we assess participants' social perception and engagement ratings of the chatbot. See Figure~\ref{fig:Likert-plots} for a visualisation of the three social perception measures. The engagement measure (Desire to Continue) is reported separately below.

For \textbf{Warmth}, there was a significant difference for \textit{Thinking Content} ($F_{2,230} = 4.85$, $p=0.0087$).
Post-hoc comparisons revealed that \textsc{Emotionally-Supportive} thinking (M = 4.71, SE = 0.16) elicited significantly higher feelings of warmth than \textsc{None} (M = 4.01, SE = 0.16, $p = 0.0078$).
There was also a significant effect for \textit{Conversational Context}, ($F(1,230) = 7.01$, $p = 0.0086$), with \textsc{Feelings} contexts (M = 4.43, SE = 0.10) rating higher than \textsc{Habit} contexts (M = 4.20, SE = 0.10).
There were no interaction effects between \textit{Thinking Content} and \textit{Conversational Context} conditions.

For \textbf{Competence}, there were no significant effects for \textit{Thinking Content} ($F_{2,230}=1.96$, $p=0.1429$), \textit{Conversational Context} ($F_{1,230}=0.03$, $p=0.8737$), or their interaction ($F_{2,230}=0.59$, $p=0.5551$).

For \textbf{Respect}, there were no significant effects for \textit{Thinking Content} ($F_{2,230}=0.95$, $p=0.3894$), \textit{Conversational Context} ($F_{1,230}=2.62$, $p=0.1066$), or their interaction ($F_{2,230}=0.29$, $p=0.7482$).

For \textbf{Desire to Continue}, there were no significant effects for \textit{Thinking Content} ($F_{2,230}=2.33$, $p=0.0997$), \textit{Conversational Context} ($F_{1,230}=0.27$, $p=0.2680$), or their interaction ($F_{2,230}=0.45$, $p=0.6353$).

\section{Qualitative Findings}

To analyse participants' responses to open-ended questions (see §~\ref{sec:qualitative_measures} for questions), we followed an inductive thematic analysis approach.
First, two members of the research team independently familiarised themselves with the qualitative data, and generated initial codes.
Next, the two research members met to discuss and clarify interpretations of the qualitative data and codes. 
Once these initial thoughts had been shared, the same two members of the team independently coded responses in detail (examples of codes include ``\textit{feeling supported}'', ``\textit{trust in suggestions}'', and ``\textit{feeling undermined}''). 
After this, the two coders discussed interpretation, before one of the coders compared codes for participant quotes, and consolidated similar codes. 
During this stage, additional discussions of potential discrepancies ensured a shared interpretation of participant quotes was reached.

When presenting participant quotes, we include both participant number and \textit{Thinking Content} condition. For example, P1[\textsc{Emotion}] would represent a participant in the \textsc{Emotionally-Supportive} condition, P2[\textsc{Expert}] an \textsc{Expertise-Supportive} participant, and P3[\textsc{None}] a \textsc{None} participant.

%Grouped codes into three broader themes of ``\textit{Direct Perceptions of Chatbot Thinking}'', ``\textit{Social Perceptions of Chatbot}'', and

% Chatbot Thinking and Intentions
% Chatbot Conversational Style and Tone
% Willingness to Follow Suggestions

\subsection{\textsc{Emotionally-Supportive} Chatbot}

%Felt ``\textit{supported}'' and ``\textit{understood}'', ``\textit{comforting}'', ``\textit{understanding}'', ``\textit{non-judgemental}''
%Several participants using phrase ``\textit{acting in my best interest}''

Within the \textsc{Emotionally-Supportive} condition, feedback primarily focused on user perceptions of the chatbot's ``\textit{thoughtful}''\footnote{Note: Where \textit{multiple} participants used the same word or short phrase to describe the chatbot, we have not provided participant numbers in the interest of space and paragraph readability.} and empathic thinking. 
%This involved participants stating that they felt ``\textit{supported}'' by the chatbot, that the chatbot was ``\textit{non-judgemental}'' and ``\textit{looking out}'' for them, and that the chatbot ``\textit{cared}'' for, wanted to ``\textit{help}'', and ``\textit{understood}'' them emotionally.
Participants often described feeling ``\textit{supported}'' by the chatbot, perceiving it as ``\textit{non-judgemental}'' and ``\textit{looking out}'' for them, and as a system that ``\textit{cared}'', wanted to ``\textit{help}'', and ``\textit{understood}'' them emotionally.
For example P30[\textsc{Emotion}] remarked: ``\textit{I felt supported by the chatbot while it was "thinking" as it seemed to me it was deliberating my circumstances and took time to answer me mindfully!}''.
%Further, PXX[\textsc{Emotion}] described the thinking as giving time for them to trust the chatbot:
Further, P73[\textsc{Emotion}] explained that the chatbot's thinking process gave them time to build trust:
\begin{quote}
    ``\textit{I felt better reading the statement as it was thinking before I read any real advice or recommendations. It allowed me to gather my thoughts and trust that it was taking its time to answer me seriously.}''
\end{quote}

On from this, several participants described appreciation for the transparency of the chatbot’s thinking, with some participants describing that it helped them feel more comfortable talking to the chatbot. For example, P215[\textsc{Emotion}] described having a sense of openness due to the chatbot’s thinking: 
\begin{quote}
    ``\textit{It was a preface that helped me understand what the chatbot was trying to convey with its suggestions and how it treated our interactions. It let me be more open with it.}''
\end{quote}

This sense of openness extended further, with some participants describing greater levels of trust in, and willingness to accept or listen to, the chatbot's suggestions. For example, P10[\textsc{Emotion}] stated: 
``\textit{I felt like when the tone was more warm it made me more willing to listen}'', while P44[\textsc{Emotion}] stated they were: ``\textit{willing to accept the suggestion based on the empathy expressed}''.

Participants in the \textsc{Emotionally-Supportive} condition also used more anthropomorphic language to discuss the chatbot, such as multiple participants comparing it to ``\textit{a friend}'' or describing it as ``\textit{real}'' and ``\textit{human}''. For example, P210[\textsc{Emotion}] stated: ``\textit{When it paused to think, it made the conversation feel more real and human. It seemed like it was considering my situation carefully, which I liked}''.

However, the \textsc{Emotionally-Supportive} thinking also primed some participants to expect suggestions that would themselves be heavily empathic in style (rather than the neutral suggestions used for all three \textit{Thinking Content} conditions). This led to negative expectancy violations with some, such as P207[\textsc{Emotion}] who stated: 
%``\textit{The AI stated that its priority was to be emotionally understanding to the user, but it didn't actually do that within its advice. It was a pretty basic response and didn't feel empathetic}''.
\begin{quote}
    ``\textit{I liked the basic end result answer of the chatbot but didn't find it to be deep enough. Especially as the "thinking" portion of its answer suggested that it would give me a lot more empathetic and personalised answer}''.
\end{quote}

When participants contrasted the sharing of \textsc{Habit}-related and \textsc{Feelings}-related problems, a similar mismatch in expectations occurred, with some participants describing what felt like an unnecessary level of empathy when sharing their \textsc{Habit}-related problems. 
For instance, P87[\textsc{Emotion}] described:
\begin{quote}
    ``\textit{The AI did understand my goal and understood that there were reasons it was difficult for me to stick to that goal. But, the subject} [Habits] \textit{was not that emotionally involved and didn't necessitate deep emotional support.}'' 
\end{quote}

Finally, some \textsc{Emotionally-Supportive} participants held utilitarian and tool-based views of chatbots, and found the idea of a chatbot expressing emotions and support through its displayed thinking as ``\textit{patronizing}'', ``\textit{weird}'', or ``\textit{inauthentic}''.
Representative of this sentiment, P137[\textsc{Emotion}] stated:
``\textit{The chatbot's response felt very artificial. It tried to express sympathy and connect with me, but it felt fake}''.
Additionally, a few participants expressed irritation and perceived lack of utility in the chatbot's thinking process. 
For instance, P1[\textsc{Emotion}] described feeling: ``\textit{irritated because it took too long}'', while another participant commented: ``\textit{I just want simple direct responses, not a pop psych opinion}''.

\subsection{\textsc{Expertise-Supportive} Chatbot}

%Much of the feedback for the Expertise-Supportive chatbot centred on what was seen as a “trustworthy”, “competent”, and “logical”, with participants describing the chatbot as understanding the content and context of their personal problems. 
Much of the feedback for the \textsc{Expertise-Supportive} chatbot centred on its perceived qualities of being ``\textit{trustworthy}'', ``\textit{competent}'', and ``\textit{logical}'', with participants noting that it appeared to understand both the content and context of their personal problems.
%This led participants to focus on the perceived quality of chatbot suggestions in their feedback, with sentiment that the \textsc{Expertise-Supportive} chatbot would produce dependable guidance, such as PXX[\textsc{Expert}] stating that the chatbot would: ``\textit{give me solid advice when it was asked}''.
This perception led many to focus on the quality and reliability of the chatbot's suggestions, expressing confidence that the \textsc{Expertise-Supportive} chatbot would provide dependable guidance. 
For instance, P217[\textsc{Expert}] commented that the chatbot would ``\textit{give me solid advice when it was asked}'', while P17[\textsc{Expert}] described the chatbot's thinking as a process of arriving at a ``\textit{logical answer}'': 
``[The chatbot] \textit{took some time to think through my problem and give a logical answer}''.
Participants also described feeling a sense of trust in the chatbot’s suggestions, with some describing a willingness to follow guidance due to the expertise-framed thinking.
As P50[\textsc{Expert}] explained: ``\textit{It increased my willingness to follow with its professional nature}''.

Similarly to the \textsc{Emotionally-Supportive} condition, some participants described valuing the transparency of the chatbot's thinking.
For instance, P92[\textsc{Expert}] noted: ``\textit{I appreciated seeing the chatbot's thought process in formulating the answer, which did make its intentions clearer}'', while P85[\textsc{Expert}] explained that:
``\textit{It definitely helped to see the "thoughts" of the chatbot. It affirmed that its information was based on what it considered reliable sources}''.
%The chatbot’s thinking also increased perceived effort on the part of the chatbot.
The chatbot's displayed thinking also contributed to a perception of increased effort on the part of the chatbot.
%These descriptions of effort assumed mostly a technical association, such as PXX[\textsc{Expert}] stating: ``\textit{I felt like it was working for me to solve my problem}'', and PXX[\textsc{Expert}] describing ``\textit{confidence}'' from ``\textit{processing}'' time: ``\textit{it was really processing before responding which gave me confidence}''.
These perceptions of effort were typically framed in technical terms, such as P168[\textsc{Expert}] noting: ``\textit{I felt like it was working for me to solve my problem}'', and  P143[\textsc{Expert}] describing a sense of ``\textit{confidence}'' arising from the chatbot's ``\textit{processing}'' time: ``\textit{it was really processing before responding which gave me confidence}''.

When describing the thinking and intentions of the \textsc{Expertise-Supportive} condition, many participants used less anthropomorphic language when referring to the chatbot, instead employing metaphors such as a ``\textit{search engine}'', ``\textit{textbook}'', or describing it as ``\textit{robotic}''.
While this was meant as a pejorative by some, it was actually seen as beneficial by others, who appreciated what they saw as a more neutral, objective, and trustworthy source of information. 
For example, P227[\textsc{Expert}] explained that:
\begin{quote}
    ``\textit{I like that the answer was not emotionally pandering and straight to the facts. I found that to be very constructive. I already have enough emotional trauma and really am looking for concise and distilled solutions from clearer heads.
    I think the fact that the response was not wrapped up in flattery or overly empathetic language, gave it more credibility.}''
\end{quote}

However, several participants also highlighted limitations of the \textsc{Expertise-Supportive} chatbot, particularly a perceived lack of empathy, social closeness, and emotional connection.
Terms such as ``\textit{cold}'', ``\textit{not personalized}'', and ``\textit{uncaring}'' were used to describe these perceptions. 
This perceived lack of empathy led some participants to feel ``\textit{dismissed}'' or ``\textit{disregarded}'' by the chatbot, and, in some cases, less willing to follow its suggestions. 
For instance, P66[\textsc{Expert}] reflected, ``\textit{I felt apathetic}'' after sharing their personal problem with the chatbot, describing a sense of diminishment and lack of care: ``\textit{I felt like it was silly and the chatbot didn’t care}''. 
Similarly, P169[\textsc{Expert}] explained: 
``\textit{The chatbot showed where I could be wrong in my loan application and told me how to fix it, but it never answered anything concerning how I was feeling. The response did not consider my emotional needs.}''

This perceived lack of empathy became particularly apparent when participants discussed their \textsc{Feelings}-related problems.
Illustrating this sentiment, P118[\textsc{Expert}] explained, ``\textit{The tone made me feel it was competent and logical, but I wouldn't necessarily want to discuss my feelings with it}''.
Several participants also described a sense of negative expectancy violation, where the \textsc{Expertise-Supportive} chatbot's visible thinking led them to anticipate more personalised or sophisticated suggestions than were ultimately delivered.
As P206[\textsc{Expert}] reflected:
\begin{quote}
    ``\textit{When the chatbot was 'thinking,' the on-screen text made me feel like it was preparing a strategic, personalized response. This actually set a higher expectation for the quality of advice. When the subsequent response was so minimal and basic, the "thinking" stage became a factor in my disappointment because the effort displayed did not match the outcome.}''
\end{quote}

% \begin{quote}
%     The chatbot understood my request, but it did not in any way acknowledge the emotional aspects I referenced, it's assistance was 100\% practical. So, while I registered that it understood my needs on a practical level, it did not do anything to help me with the emotional aspects of what I expressed need for help with.
% \end{quote}

\subsection{\textsc{None} Condition Chatbot}

Finally, although the \textsc{None} condition chatbot did not display thinking to participants, there was some overlap in sentiment with the \textsc{Expertise-Supportive} condition regarding what was seen as a lack of empathy.
However, in contrast to the \textsc{Expertise-Supportive} condition, this absence of empathy led to greater expressions of unwillingness to follow the chatbot's suggestions, alongside a recurring sentiment that the interaction was ``\textit{boring}'' and lacked engagement.
Illustrative of this, P150[\textsc{None}] reflected: 
``\textit{It gave good advice but lacked the emotional weight of it so I probably won't follow its advice mainly out of boredom and carelessness rather than that I felt it was a wrong thing to do}''. 

In addition to this sense of boredom, there was also a perception that the chatbot demonstrated minimal effort.
For example, P39[\textsc{None}] lamented:
\begin{quote}
    ``\textit{it felt like the AI put minimal effort into its response, just telling me to enjoy the present and talk to my partner. To be fair though, I suppose there isn't much more advice to give.}''
\end{quote}

Unique to the \textsc{None} condition feedback, were perceptions that the chatbot demonstrated ``\textit{no interest}'' in the user themselves. For instance, P107[\textsc{None}] explained:
\begin{quote}
    ``\textit{It expressed no interest in me at all. No attempt at understanding or acknowledging my situation.}''
\end{quote}

%Beyond this, the \textsc{None} condition chatbot was seen as focusing on provision of information (e.g., PXX[\textsc{None}]: ``\textit{I felt the chatbot was just providing stock answers and does not provide anything beyond generic answers.}''), and less anthropomorphic language was used to describe the chatbot (e.g., ``\textit{robotic}'', ``\textit{machine}''). 
%This led to repeated sentiment that participants would rather use a search engine for their help-seeking than talk to the \textsc{None} condition chatbot. 
%For example, PXX[\textsc{None}] proffered:

Beyond this, the \textsc{None} condition chatbot was viewed primarily as an informational tool.
Participants described it as providing ``\textit{stock}'' or ``\textit{generic}'' responses (e.g., P35[\textsc{None}]: ``\textit{I felt the chatbot was just providing stock answers and does not provide anything beyond generic answers.}''), and used less anthropomorphic language overall (e.g., ``\textit{robotic}'', ``\textit{machine}''). 
This led to a recurring sentiment that participants would rather use a search engine for help-seeking than engage with the chatbot. 
As one participant observed:
\begin{quote}
    ``\textit{This chatbot wasn't emotionally responsive or empathetic at all. It just made a few suggestions that might be totally unhelpful for me personally} [...]. \textit{This is no better than I could get from a search engine AI.}''
\end{quote}

%Large amounts of sentiment was shared between the \textsc{None} and \textsc{Expertise-Supportive} conditions, regarding what was seen as language devoid of empathy.
%Participants in both the \textsc{None} and \textsc{Expertise-Supportive} conditions described feeling ``\textit{undermined}'' or ``\textit{dismissed}'' by what was seen as language devoid of empathy.
%Some participants described feeling ``\textit{undermined}'', ``\textit{dismissed}'', or ``\textit{disregarded}'', such as PXX[\textsc{None}] stating: ``\textit{the chatbot totally undermined my situation and I didn't really feel any type of sympathy for my situation at all}''; and PXX[\textsc{Expert}] describing: ``\textit{The chatbot gave me a very objective answer and that made me feel significantly disregarded}''.
%For example, PXX[\textsc{None}] stated: 
%``\textit{the chatbot totally undermined my situation and I didn't really feel any type of sympathy for my situation at all}''.
%Similarly, PXX[\textsc{Exp}] stated: ``\textit{The chatbot gave me a very objective answer and that made me feel significantly disregarded}''.

Taken together, these findings suggest that visible thinking served as an important social cue for perceived effort, competence, and empathy. 
In its absence, as in the \textsc{None} condition, participants described interactions as impersonal, less engaging, and ultimately less motivating to act upon.

\section{Discussion}

Conversational agents can offer users a glimpse into their internal reasoning processes. 
In practice, this narration occurs while the system is concurrently generating a response using its underlying mechanisms, which may or may not be related to the displayed ``thinking'' process.
The displayed ``thinking'' or activity cue may serve as a communicative layer rather than a literal representation of computation. 
Thus, it may also shape user expectations and perceptions of the chatbot and its output.
%We explored how people may perceive a conversational agent's reasoning and thoughts, as well as how such reasoning could be designed to make users feel more comfortable.
We explored how the visibility and framing of a chatbot's thinking affects user perceptions (such as feelings of empathy, warmth, and competence) in both a \textsc{Emotionally-Supportive} thinking, \textsc{Expertise-Supportive} thinking, and \textsc{None}, where no thinking was displayed to users.
Users shared both \textsc{Habit}-related and \textsc{Feelings}-related personal problems with the chatbot.

\subsection{The Impact of Visible Chatbot Thinking}

Our findings show that both the visibility and framing of a chatbot's `thinking' shape user perceptions and expectations during help-seeking conversations.
Specifically, the chatbot with \textsc{Emotionally-Supportive} thinking was rated as more emotionally responsive, empathic, and warm.
Participants in this condition frequently described the chatbot as ``\textit{thoughtful}'', ``\textit{supportive}'', and ``\textit{caring}'', noting that the visible thinking helped them feel emotionally understood and fostered a sense of openness and comfort when discussing personal problems.
However, \textsc{Emotionally-Supportive} thinking also primed some participants to expect deeply empathic responses from the chatbot, creating negative expectancy violations~\cite{BURGOON201624} when those expectations were unmet.
%leading to disappointment when this was seen as a negative expectancy violation. 
Additionally, some participants describe the chatbot's thinking as ``\textit{patronizing}'' or ``\textit{inauthentic}''.
%described how the warm and empathic perceptions of the chatbot encouraged feelings of openness and comfort
In contrast, the \textsc{Expertise-Supportive} chatbot was rated as possessing greater levels of trust and understanding (compared to the \textsc{None} condition).
Participants described the chatbot as ``\textit{logical}'', ``\textit{professional}'', and ``\textit{reliable}'', and appreciated both the transparency and perceived effort conveyed through its visible thinking.
However, participants also characterised the \textsc{Expertise-Supportive} chatbot as ``\textit{cold}'' and ``\textit{uncaring}'' leading some to express reluctance in sharing personally emotional experiences with the chatbot.
Additionally, some reported frustration when the depth implied by the chatbot's `\textit{thinking}' was not reflected in the substance of its eventual reply.
Finally, the \textsc{None} condition (without any visible thinking) was seen as lacking empathy or interest in the user, and described as impersonal, less engaging, and less motivating to act upon.

% This work sits in relation with prior work that explores adapting social cues related to a chatbot's framing~\cite{cox2025ephemerality} or metaphor~\cite{khadpe2020conceptual,10.1145/3640794.3665535,10.1145/3609326}.
% Our findings indicate that, beyond the manipulation of verbal cues present in a chatbot's final utterances, the verbal cues present in the `thinking' of chatbots can shape and prime user perceptions and expectations.
% As `thinking' and `reasoning' becomes ever more present in conversational agents, reflecting on the design and content of such systems becomes more important, with designers needing to think beyond ``thinking'' as a means of explainability and additionally considering the social impact on users.

% The \textsc{Emotionally-Supportive} condition follows findings of prior work that the use of empathic language can encourage...
% Additionally, perceptions of this language as being condescending mirror recent work that the use of politeness strategies in conversational agents can sometimes backfire and appear insincere~\cite{bowman2024exploring}.

This work sits in relation to prior research on the adaptation of social cues in conversational agents, such as manipulations of framing~\cite{cox2025ephemerality} or metaphor~\cite{khadpe2020conceptual,10.1145/3640794.3665535,10.1145/3609326}. Our results extend this literature by indicating that, beyond the verbal cues present in a chatbot’s final utterances, the language embedded in a chatbot's visible `\textit{thinking}' can also prime user expectations and shape perceptions of empathy, trust, and effort. As `\textit{thinking}' and `reasoning' displays become more commonplace in conversational agents, designers must consider their social as well as explanatory impact, reflecting on how such `\textit{thinking}' conveys intention and personality to users.
Considering the impact of `\textit{thinking}' displays is increasingly pertinent given the rising use of commercially available LLMs for help-seeking~\cite{jung2025ve}.

% The \textsc{Emotionally-Supportive} condition aligns with prior findings that empathic language can foster greater feelings of warmth, emotional connection, and comfort in disclosing~\cite{skjuve2022longitudinal,10.1145/3313831.3376175}.
% At the same time, perceptions of this language as condescending by some, demonstrates an unintended effect (similarly to prior CA research finding politeness strategies can be seen as inauthentic~\cite{bowman2024exploring}, and that conversational styles may unintentionally make users feel \textit{more} rather than less stressed~\cite{10.1093/jcmc/zmab005}).
%mirror recent observations that politeness strategies can backfire when perceived as inauthentic~\cite{bowman2024exploring}. 
%Similarly, the \textsc{Expertise-Supportive} condition highlights that visible thinking can enhance perceived competence and trust, mirroring prior work exploring the framing of CAs as experts~\cite{10.1145/3613905.3650749}.
%However, \textsc{Expertise-Supportive} thinking inadvertently distanced users in more affective contexts.
%if the thinking is not framed as sufficiently empathic. 
%Together, these patterns reinforce that the design of visible reasoning should balance cognitive transparency with social attunement.
The \textsc{Emotionally-Supportive} condition aligns with prior findings that empathic language can foster greater feelings of warmth, emotional connection, and comfort in disclosing~\cite{skjuve2022longitudinal,10.1145/3313831.3376175}.
At the same time, perceptions of this language as condescending by some participants demonstrate an unintended effect, echoing prior CA research showing that politeness strategies can be perceived as inauthentic~\cite{bowman2024exploring}, and that conversational styles may unintentionally increase stress rather than reduce it~\cite{10.1093/jcmc/zmab005}.
Similarly, the \textsc{Expertise-Supportive} condition highlights that visible thinking can enhance perceived competence and trust, mirroring prior work exploring the framing of CAs as experts~\cite{10.1145/3613905.3650749}.
However, \textsc{Expertise-Supportive} thinking inadvertently distanced users in more affective contexts.
Additionally, users described their own preferences and beliefs in chatbots as affecting their perceptions of \textit{thinking} displays (such as those with a utilitarian view of chatbots appreciating the \textsc{Expertise-Supportive} condition).
This reflects broader work within conversational agents that argues against a one-size-fits-all approach, and motivates for personalised or user customisable conversational styles~\cite{10.1145/3715336.3735795,10.1145/3706598.3713453}.
Together, these findings suggest that supportive and expert framings require careful calibration, as they can simultaneously increase perceived warmth or competence while risking inauthenticity or emotional distance.

%However, both of the thinking conditions also caused expectancy violations when the primed content 
%Further, both of the visible thinking conditions caused negative expectancy violations where primed expectations from the chatbot's thinking did not align with user expectations once the chatbot's final suggestion was presented (with such expectancy violations having the potential to cause lessened feelings of connectedness, competence, and being understood~\cite{BURGOON201624,10.1145/3290605.3300641}).
Both visible thinking conditions also produced negative expectancy violations when the chatbot's final reply failed to meet the expectations established by its displayed thinking, which may lessen feelings of connectedness, competence, and understanding~\cite{BURGOON201624,10.1145/3290605.3300641,10.1145/3717511.3747079}.
On from this, in the case of both \textsc{Emotionally-Supportive} and \textsc{Expertise-Supportive} conditions, some participants described an incongruence between the values being framed by the chatbot and the type of problem (\textsc{Habit}-related or \textsc{Feelings}-related) being disclosed.
Such negative consequences of both priming and conversational context highlight the importance of designing verbal cues for `thinking' that both ensure users feel supported and understood, while also possessing appropriate alignment with the chatbot's intended role and the user’s situational needs. Designing visible thinking that adapts to conversational context may therefore help prevent expectancy violations and support more authentic, meaningful engagement.

Participants in both visible thinking conditions also reflected on the temporal aspects of thinking displays.
Common sentiment among participants was that \textit{thinking} time reflected an element of effort on the part of the CA. 
Here \textsc{Emotionally-Supportive} participants described feeling that the agent was pausing to think about them and to understand their emotional needs; while \textsc{Expertise-Supportive} participants described a more mechanical process, perceiving processing time as effort to generate a high quality suggestion.
These feelings also impacted trust and intended compliance towards the chatbot's suggestions.
Matching prior XAI literature in decision-making contexts, the thinking displays of the \textsc{Expertise-Supportive} chatbot led participants to describe more trust and acceptance of the chatbot~\cite{he2025conversationalxai,joshi2024explainability}.
This also reflects recent research within conversational user interfaces that (in contrast to prior research aiming to minimise response latency) the presence of delay in conversational responses can actually improve user perceptions in some contexts~\cite{kim2025fromseconds}.

\subsection{Limitations and Future Work}

% This study investigated single-turn interactions with conversational agents. While the use of single-turn interactions has been adopted in related work~\cite{10.1145/3613904.3642135}, reduced ecological validity (compared to ongoing dialogue) should be highlighted.
% However, the use of a single-turn interaction, both allowed for greater experimental control (ensuring more consistent and comparable experience between participants), and also mirrors the increasing use of conversational agents such as ChatGPT, where people may ask for solutions to their personal problems~\cite{jung2025ve}.

This study examined single-turn interactions with a conversational agent. While single-turn designs are common in related work~\cite{10.1145/3613904.3642135}, they reduce ecological validity relative to multi-turn, ongoing dialogue. However, this design enabled greater experimental control and more consistent experiences across participants, and also reflects common usage patterns in systems such as ChatGPT, where users may seek solutions to personal problems in brief interactions~\cite{jung2025ve}.

%Further, the final chatbot suggestions were all generated using the same LLM prompting independent of \textit{Thinking Content} condition. This mirrors prior work where underlying chatbot behaviour is fixed between conditions (e.g.,~\cite{khadpe2020conceptual}), and allowed to (within the non-deterministic confines of LLM-based research) isolate the manipulation's effect to only the content of the chatbot's thinking. Future work could additionally investigate manipulations of both the chatbot's thinking and output.
Second, the chatbot's final suggestions were generated using the same prompting across \textit{Thinking Content} conditions. This mirrors prior work that holds underlying system behaviour constant while varying framing or presentation~\cite{khadpe2020conceptual}, and helped isolate the effects of visible thinking content within the constraints of non-deterministic LLM outputs. Future work could vary both the agent's thinking and its final responses to examine how these factors interact.

% Additionally, the thinking time was fixed to five seconds (i.e., the chatbot's thinking was shown on screen for five seconds, before the chatbot suggestion was shown to users) for all interactions.
% Another consideration is the thinking time and how this affects people's perceptions and decision-making practices. 
% Within our study, thinking time was fixed to five seconds (i.e., the chatbot's reasoning was shown on screen for five seconds, before it's main response was shown to users). 
% Recent work has also investigated the impact of chatbot response time on...~\cite{kim2025fromseconds}.
% %Waiting time in image generation (some people value as part of creative process~\cite{10.1145/3706599.3719725}
% Phrasing of the thinking content (use more human like tone or vocabulary)
% Length of the thinking content 
Another consideration concerns the presentation and duration of the chatbot's visible thinking. 
The thinking display was fixed to five seconds for all interactions. After this delay, a ``\textit{Done}'' indicator appeared next to the thinking to show completion, and the chatbot's final suggestion was presented below in a separate message bubble. 
While this consistency supported comparison across conditions, future studies could vary timing, pacing, and display dynamics of thinking, building on work examining response latency and its effects on user perception~\cite{kim2025fromseconds}.

Lastly, we explored within the domain of people sharing personal problems and help-seeking, a domain where emotional support may be particularly salient. 
%Future studies could investigate perceptions in other domains and use cases, as well as investigations of the use of different roles and formats of thinking or reasoning text.
Future studies could investigate whether these effects generalise to other domains and use cases, and explore alternative roles and formats for visible thinking or reasoning text.

\section{Conclusion}
% In this study we investigated how users perceive chatbot ``thinking'' indicators during the conversational interaction. 
% We conducted a mixed-design experiment in which participants engaged in chatbot discussions on two distinct topics.
% During response generation, the chatbot displayed one of two textual cues: \textit{Emotionally-Supportive} or \textit{Expertise-Supportive}, or a baseline (no cue).
% The former emphasised acknowledgement of users’ emotions, while the latter signalled an analytical, fact-oriented stance.
% This allowed us to explore how different signalling styles during the chatbot’s visible response generation phase influence users' perceptions about the chatbot.
% We find statistically significant differences between the conditions, e.g. in emotional responsiveness, trustworthiness, and perceived empathy.
% Our study is one of the very first in HCI to explore the effects of these glimpses on a chatbot's internal reasoning process, showing how they affect users' immediate perceptions.
% The results indicate how the design of thinking cues should be taken into account when building e.g. emotionally intelligent chatbots.

In this study, we investigated how users perceive a conversational agent when it displays visible \textit{thinking} during response generation. 
We conducted a mixed-design experiment in which participants discussed both \textsc{Habit}-related and \textsc{Feelings}-related personal problems with a chatbot. 
Participants were assigned to one of three \textit{Thinking Content} conditions: \textsc{None}, \textsc{Emotionally-Supportive}, or \textsc{Expertise-Supportive}. 
By varying whether the agent showed no thinking or presented value-oriented reflections prior to responding, we examined how the presence and framing of visible thinking shape users’ impressions of the chatbot. 
We found significant differences across conditions in key perceptions including emotional responsiveness, understanding and trust, and perceived empathy. 
Our work contributes early evidence on how visible \textit{thinking} displays influence user perceptions in emotionally sensitive, help-seeking conversations.
Together, our findings highlight that visible \textit{thinking} functions as a form of chatbot self-presentation that can meaningfully shape user experience in help-seeking contexts, with implications for designing conversational agents that communicate intentions appropriately in emotionally sensitive interactions.

% \section{GenAI Usage Disclosure}
% ChatGPT was employed to enhance writing quality across sections of the paper.

% \begin{acks}
% \end{acks}

\bibliographystyle{ACM-Reference-Format}
\bibliography{sample-base}

\newpage
\appendix

\section{Example \textsc{Habit}-related Problem Chatbot and User Utterances}

\begin{table}[h]
  \caption{Examples of user utterances (\textsc{Habit}-related problems) and chatbot utterances by \textit{Thinking Content} conditions.}
  \label{tab:utterances_conditions-habits}
  \centering
  \footnotesize
  \renewcommand{\arraystretch}{1.4} % increase vertical spacing
  %\begin{tabular}{p{2.5cm} p{4.5cm} p{4.5cm} p{4.5cm}} 
  \begin{tabular}{p{0.15\linewidth} p{0.17\linewidth} p{0.29\linewidth} p{0.29\linewidth}} 
    \toprule
     &  Problem Shared by User & Thinking Utterance & Chatbot Suggestion \\
    \midrule
    \textsc{Emotionally-Supportive} & 
    ``\textit{I would like to drink more water and less soda.}'' &
    ``\textit{I aim to be emotionally supportive by acknowledging the user's healthy intention to drink more water, offering encouragement and validation for this positive change, and ensuring that the user feels empowered and non-judged in their journey towards better hydration habits.}'' &
    ``\textit{Try carrying a reusable water bottle with you to stay hydrated and slowly replace soda with flavored or sparkling water to help adjust your taste preference.}'' \\

    \textsc{Expertise-Supportive} & 
    ``\textit{I am not a very healthy eater because I just don't like to eat. So most days I barely eat anything because I only eat when I think about it and I just don't think about eating. What are some ways I can start being healthier about eating?}'' &
    ``\textit{My aim is to equip the user with reliable and expert-backed strategies for developing healthier eating habits, providing well-researched information and guidance from nutrition experts to ensure the user receives accurate and practical advice grounded in scientific knowledge.}'' &
    ``\textit{Try setting a regular meal schedule with alarms as reminders, focusing on balanced meals with small portions of nutritious foods you enjoy, and gradually incorporating healthy snacks throughout the day to stimulate your appetite.}'' \\   

    \textsc{None} & 
    ``\textit{I'd like to learn a new skill.}'' &
    --- &
    ``\textit{Consider using online platforms like Coursera or Udemy to explore courses on skills of interest, such as coding, graphic design, or language learning.}'' \\
    \bottomrule
  \end{tabular}
\end{table}

\end{document}